%Paper: hep-th/9409021
%From: a.tseytlin@ic.ac.uk
%Date: Sat, 3 Sep 94 21:24:14 bst
%Date (revised): Fri, 30 Sep 94 00:24:27 bst
%Date (revised): Fri, 30 Sep 94 00:30:34 bst

\input harvmac
\noblackbox

\def \hi {{\hat i}}
\def \hj {{\hat j}}\def \hk {{\hat k}}\def \hu {{\hat u}}\def \hv {{\hat v}}

\def \FM {$F$-model\  }
\def \KM {$K$-model\  }

\def \FMS{$F$-models\  }

\def \eq#1 {\eqno {(#1)}}

\def \ra {\rightarrow}
\def\np {  Nucl. Phys. }
\def \pl { Phys. Lett. }
\def \mpl { Mod. Phys. Lett. }
\def \prl { Phys. Rev. Lett. }
\def \pr  { Phys. Rev. }

\def \ap  { Ann. Phys. }
\def \cmp { Commun. Math. Phys. }
\def \ijmp { Int. J. Mod. Phys. }

%%%%%%%%%%%%%%%%%%%%%%%%%%%%%%%%%%%%%%%%%%%%%%%%%%%%%%%%%%%%%%
\def\k{\kappa}
\def\r{\rho}
\def\a{\alpha}
\def\b{\beta}

\def\g{\gamma}
\def\G{\Gamma}
\def\d{\delta}

\def\e{\epsilon}

\def\p{\phi}

\def\m{\mu}
\def\n{\nu}
\def\om{\omega}

\def\l{\lambda}

\def\s{\sigma}

\def\cA{{\cal A}}
\def \sm {$\s$-model\ }

\def \bd {\bar \del}

\def \A { {\bar A} }

\def \ha {{1\over 2}}

\def \ov {\over}

\def\const{{\rm const}}
\def \p {\phi}
\def \vp {\varphi}
%%%%%%%%%%%%%%%%%%%%%%%%%%%%%%

\def \sms {$\s$-models\ }

\def \bd {\bar \del}

\def \tt {{\tilde \t}}
\def \G {\Gamma}

\def \ra {\rightarrow}

\def \a {\alpha}
\def \b {\beta}
\def \chi {\chi}

\def \ln {{\rm \ ln \  }}

\def \l {\lambda}
\def \p {\phi}

\def \m {\mu }
\def \n {\nu}
\def \ep {\epsilon}
\def\g {\gamma}
\def \r {\rho}
\def \k {\kappa }
\def \d {\delta}

\def \s {\sigma}

\def \fourth {{\textstyle{1\over 4}}}

\def \e#1 {{{\rm e}^{#1}}}
\def \const {{\rm const }}

\def \eq#1 {\eqno {(#1)}}
%%%%%%%%%%%%%%%%%%%%%%%%%%%%%%%%%%%%%%%%%%%%
\def \sm {$\s$-model\ }

\def \bd  {{ \bar \del }}

\def \bd  { \bar \del }

\def \A { \bar A}

%%%%%%%%%%%%%%%%%%%%%%%%%%%%%%%%%%%%%%%
\def \H {{\cal H}}

\def \p {\phi}
\def \ep {\epsilon}
\def \s {\sigma}

\def \r {\rho}
\def \d {\delta}
\def \l {\lambda}
\def \m {\mu}

\def \g {\gamma}
\def \n {\nu}

\def \fourth {{1\over 4}}

\def \e#1 {{{\rm e}^{#1}}}
\def \const {{\rm const }}\def \vp {\varphi}

\def \H {{\cal H}}

%%%%%%%%%%%%%%%%%%%
\def \m {\mu}  

\def \ep {\epsilon}

\def \ra {\rightarrow}

\def \const {{\rm const} }

\def \eq#1 {\eqno{(#1)}}
\def \e {\rm e}
\def \ra {\rightarrow }
\def \e#1 {{\rm e}^{#1}}

\def \ln { {\rm ln } }
\def \sin {\ {\rm sin} \ }
\def \cos {\ {\rm cos} \ }

\def \l {\lambda}
\def \p {\phi}
\def \vp {\varphi}
\def  \g {\gamma}

\def \r {\rho}

\def\({\left (}
\def\){\right )}
\def\[{\left [}
\def\]{\right ]}

\def \cA {{\cal A}}
 \def \cB {{\cal B}}

\def \hG {\hat \Gamma}

\def \tt{\theta}
\def \bt{{\bar \theta}}
\def \A {\tilde A}
\def \K {\tilde K}

\def\np {  Nucl. Phys. }
\def \pl { Phys. Lett. }
\def \mpl { Mod. Phys. Lett. }
\def \prl { Phys. Rev. Lett. }
\def \pr  { Phys. Rev. }
\def \cqg { Class. Quantum Grav.}
\def \jmp { Journ. Math. Phys. }
\def\ap { Ann. Phys. }

\baselineskip8pt
\Title{\vbox
{\baselineskip 6pt{\hbox{  }}{\hbox
{Imperial/TP/93-94/54 }}{\hbox {UCSBTH-94-31}}{\hbox{hep-th/9409021}} } }
{\vbox{\centerline {  A new class of exact solutions in string theory }
\centerline { }
}}
\vskip 37 true pt
\centerline  { {Gary T. Horowitz\footnote {$^*$} {e-mail address:
gary@cosmic.physics.ucsb.edu} }}

 \smallskip \smallskip
\centerline {\it Physics
Department }
\smallskip

\centerline{\it  University of California, Santa Barbara, CA 93106, USA}
\medskip
\centerline {and}
\medskip
\centerline{   A.A. Tseytlin\footnote{$^{\star}$}{\baselineskip8pt
On leave  from Lebedev  Physics
Institute, Moscow, Russia.} }

\smallskip\smallskip
\centerline {\it  Theoretical Physics Group, Blackett Laboratory}

\centerline {\it  Imperial College,  London SW7 2BZ, U.K. }
\bigskip
\centerline {\bf Abstract}
\medskip
\baselineskip11pt
\noindent
We prove that a large class of leading order string solutions which
generalize both the plane-wave and fundamental string backgrounds are, in fact,
exact solutions to all orders in $\a'$. These include, in particular,
the traveling
waves along the fundamental string.
The key features of these solutions
are a null symmetry and a chiral coupling of the string to the background.
Using dimensional reduction, one finds that
the extremal electric dilatonic black holes and their
recently discovered generalizations with  NUT charge and rotation
are also exact solutions. We show that  our  bosonic solutions  are also  exact
solutions of  the   heterotic string theory with no extra gauge field
background.

\Date {September  1994}

%\draftmode
\noblackbox
%\baselineskip 20pt plus 2pt minus 2pt
%%%%%%%%%%%%%%%%%%%%%%%%%%%%%%%%%%%%%%%%%%%%%%%%%%%%%%%%%%%%%%%%

\vfill\eject

\def \lr { \lref}

\gdef \jnl#1, #2, #3, 1#4#5#6{ {\sl #1~}{\bf #2} (1#4#5#6) #3}

\lr \mans { P. Mansfield and J. Miramontes, \jnl \pl, B199, 224, 1988;
A. Tseytlin, \jnl \pl, B208, 228, 1988; \jnl \pl, B223, 165, 1989.}
\lr \horts { G. Horowitz and A. Tseytlin, ``On exact solutions and
singularities in string theory", preprint  Imperial/TP/93-94/38,
hep-th/9406067, to
appear in Phys. Rev. D.}

\lr \dgkt{F. Dowker, J. Gauntlett, D. Kastor and J. Traschen, \jnl \pr,
D49, 2909, 1994.}

\lr \kalmor{R. Kallosh and A. Morozov,  \jnl \ijmp,  A3, 1943, 1988.}

\lr \ghrw{J. Gauntlett, J. Harvey, M. Robinson and D. Waldram,
\jnl \np, B411, 461, 1994.}
\lr \garf{D. Garfinkle, \jnl \pr, D46, 4286, 1992.}

\lr \onofri { V. Fateev, E. Onofri and Al. Zamolodchikov, \jnl \np, B406,
 521, 1993.}
\lr \alvf {L. Alvarez-G\'aume,  D. Freedman and  S. Mukhi, \jnl \ap, 134, 85,
1981; B. Fridling and A. van de Ven, \jnl \np, B268, 719, 1986.}

\lref \tspl {A. Tseytlin, \jnl \pl, B317, 559, 1993.}
\lref \tssfet { K. Sfetsos and A.  Tseytlin, \jnl  \pr, D49, 2933, 1994.}

\lref \klts {C. Klim\v c\'\i k  and A. Tseytlin, \jnl \np, B424, 71, 1994.}

\lr \sfexac {K. Sfetsos,  \jnl \np, B389, 424,  1993.}

\lr \tsmac{A. Tseytlin, \jnl \pl,  B251, 530, 1990.}

\lr \cakh{C. Callan and R. Khuri, \jnl \pl, B261, 363, 1991;
R. Khuri, \jnl \np, B403, 335, 1993.}
\lr \dgt{M. Duff, G. Gibbons and P. Townsend, ``Macroscopic superstrings
as interpolating solitons", DAMTP/R-93/5, hep-th/9405124.}
\lr \call{C. Callan, D. Friedan, E. Martinec and  M. Perry, \jnl \np, B262,
593, 1985.}
\lr \frts {E.  Fradkin  and A. Tseytlin, \jnl \pl, B158, 316, 1985;
\jnl \np, B261, 1, 1985.}
\lr \tsred  {A. Tseytlin, \jnl  \pl, B176, 92, 1986; \jnl  \np, B276, 391,
 1986.}
%\lr \grwi   { D. Gross and E. Witten, \jnl \np, B277, 1, 1986.}

\lr \gps {S.  Giddings, J. Polchinski and A. Strominger, \jnl  \pr,  D48,
 5784, 1993. }

\lr \tsppl  {A. Tseytlin, \jnl   \pl,  B208, 221, 1988.}
\lr\rabi  {S. Elitzur, A. Forge and E. Rabinovici, \jnl \np, B359, 581, 1991;
 G. Mandal, A. Sengupta and S. Wadia, \jnl \mpl,  A6, 1685, 1991. }
 \lr \witt{ E. Witten, \jnl \pr, D44, 314, 1991. }
 \lr \dvv { R. Dijkgraaf, H. Verlinde and E. Verlinde, \jnl \np, B371,
269, 1992.}
\lr \hoho { J. Horne and G.  Horowitz, \jnl \np, B368, 444, 1992. }
\lr \horwel{G. Horowitz and D. Welch, \jnl \prl, 71, 328, 1993;
N. Kaloper,  \jnl \pr,  D48, 2598, 1993. }
\lr \host{ G. Horowitz and A. Steif,  \jnl \prl, 64, 260, 1990; \jnl \pr,
D42, 1950, 1990;  G. Horowitz, in: {\it
 Strings '90}, (eds. R Arnowitt et. al.)
 World Scientific (1991).}
\lr \busch {T.  Buscher, \jnl \pl, B194, 59, 1987; \jnl \pl,
 B201, 466, 1988.}
\lr \kallosh {E. Bergshoeff, I. Entrop and R. Kallosh,
\jnl \pr, D49, 6663, 1994.}

\lr \tsmpl {A. Tseytlin, \jnl  \mpl, A6, 1721, 1991.}
\lr \kltspl { C. Klim\v c\'\i k and A. Tseytlin, \jnl \pl, B323, 305, 1994.}
\lr \shwts { A. Schwarz and A. Tseytlin, \jnl \np, B399, 691, 1993.}
\lr \callnts { C. Callan and Z. Gan, \jnl  \np, B272, 647, 1986;  A. Tseytlin,
\jnl \pl,
B178, 34, 1986.}

\lr \guv { R. G\"uven, \jnl \pl, B191, 275, 1987;
 D. Amati and C. Klim\v c\'\i k,
\jnl \pl, B219, 443, 1989; G. Horowitz and A. Steif,  \jnl \prl, 64, 260,
1990.}
 \lr\rudd{R. Rudd, \jnl \np, B352, 489, 1991.}
\lr \desa{ H. de Vega and N. Sanchez, \jnl
\pr, D45, 2783, 1992; \jnl \cqg, 10, 2007, 1993.}
\lr \desas{ H. de Vega and N. Sanchez, \jnl
\pl, B244,  215, 1990.}
\lref \tsnul { A. Tseytlin, \jnl \np, B390, 153, 1993.}

\lref \dunu { G. Horowitz and A. Steif, \pl B250 (1990) 49;
 E. Smith and J. Polchinski, \pl B263 (1991) 59. }

\lr \gauged {I. Bars and K. Sfetsos, \jnl  \mpl, A7, 1091, 1992;
 P. Horava, \jnl \pl,
B278, 101, 1992; P. Ginsparg and F. Quevedo, \jnl \np, B385, 527, 1992. }
\lr \bsfet {I. Bars and K. Sfetsos, \jnl \pr, D46, 4510, 1992; \jnl \pr,
 D48, 844, 1993. }
\lr \tsnp{ A. Tseytlin, \jnl \np, B399, 601, 1993;  \jnl \np, B411, 509, 1994.}
\lr \gibb{A. Dabholkar, G. Gibbons, J. Harvey and F. Ruiz Ruiz, \jnl \np, B340,
33, 1990.}
\lr \hhs{J. Horne, G. Horowitz and A. Steif, \jnl \prl, 68, 568, 1992;
G. Horowitz,
in {\sl String Theory and
Quantum Gravity '92}, eds. J. Harvey et al. (World Scientific, 1993)}
\lr \polypolchnats {  }
\lr \jack {I. Jack, D.  Jones and J. Panvel, \jnl \np, B393, 95, 1993.}
\lr \mettstwo {R.  Metsaev and A. Tseytlin, \jnl \pl, B185, 52, 1987.}
\lr \banks {T. Banks, M. Dine, H. Dijkstra and W. Fischler, \jnl \pl, B212,
45, 1988.}
\lr \horstr{G. Horowitz and A. Strominger, \jnl \np, B360, 197, 1991.}

\lr \givkir {A.  Giveon and E. Kiritsis, \jnl \np, B411, 487, 1994.  }

\lr \jac{I. Jack and D. Jones, \jnl \pl, B200, 453, 1988.}
\lr \metts{R. Metsaev and A. Tseytlin, \jnl \np,  B293, 385, 1987.  }

\lr \gib { G. Gibbons, \jnl \np, B207, 337, 1982. }

\lr \gim { G. Gibbons and K. Maeda,  \jnl \np, B298, 741, 1988. }
\lr\gar { D. Garfinkle, G. Horowitz and A. Strominger, \jnl \pr,  D43, 3140,
1991; {\bf D45} (1992) 3888(E).}

\lr\kall { R. Kallosh, D. Kastor, T. Ort\'in   and T. Torma, ``Supersymmetry
and stationary solutions in dilaton-axion gravity",
SU-ITP-94-12, hep-th/9406059. }

\lr\jons { C. Johnson and R.  Myers, ``Taub-NUT dyons in heterotic string
theory",
 IASSNS-HEP-94/50, hep-th/9406069. }

\lr\duval  { C. Duval, Z. Horvath and P.A. Horvathy, \jnl \pl,  B313, 10,
1993.}

\lr \jon {C. Johnson,  ``Exact models of extremal dyonic 4D black hole
solutions of heterotic string theory", IASSNS-HEP-94/20,
hep-th/9403192.}

\lr\bergsh { E. Bergshoeff, R. Kallosh and T. Ort\'in, \jnl \pr,  D47, 5444,
1993.}
\lr \wald {D. Waldram, \jnl \pr, D47, 2528, 1993.}
\lr \sfts { K. Sfetsos and A. Tseytlin,  `Four Dimensional Plane Wave String
Solutions with Coset CFT Description',  THU-94/08, hep-th/9404063. }

\lr \bergkal { E. Bergshoeff, R. Kallosh and T. Ort\'in, `Black-hole-wave
duality in string theory',  SU-ITP-94-11, hep-th/9406009. }
\lr\sen{A. Sen, \jnl \np, B388, 457, 1992. }

\lr \huwi{ C. Hull and  E. Witten, \jnl \pl, B160, 398, 1985. }

\lr \mukhi{S. Mukhi, \jnl \pl, B162, 345, 1985.}

\lr\iwp{W. Israel and G. Wilson, \jnl \jmp, 13, 865, 1972;
Z. Perj\'es, \jnl \prl, 27, 1668, 1971.}

\lr\maha{J. Maharana and J. Schwarz, \jnl \np, B390, 3, 1993.}
\lr \hrt  { G. Horowitz and A. Tseytlin, ``Extremal black holes as exact string
solutions", preprint  Imperial/TP/93-94/51, UCSBTH-94-24, hep-th/9408040.}
\lr \brink { H. Brinkmann, \jnl {\it Math. Ann.}, 94, 119,  1925.}

\lr \bergkal { E. Bergshoeff, R. Kallosh and T. Ort\'in, ``Black-hole-wave
duality in string theory",  SU-ITP-94-11, hep-th/9406009. }

\lr \kalor { R. Kallosh and T. Ort\'in, ``Exact $SU(2)\times U(1)$ stringy
black holes", SU-ITP-94-27, hep-th/9409060. }

\lr\sen{A. Sen, \jnl \np, B388, 457, 1992. }

\lr \senn {A. Sen, \jnl \pr, D32, 2102, 1985.}

\lr \huwi{ C. Hull and  E. Witten, \jnl \pl, B160, 398, 1985. }

\lr \mukhi{S. Mukhi, \jnl \pl, B162, 345, 1985.}

\lr\iwp{W. Israel and G. Wilson, \jnl \jmp, 13, 865, 1972;
Z. Perj\'es, \jnl \prl, 27, 1668, 1971.}

\lr\maha{J. Maharana and J. Schwarz, \jnl \np, B390, 3, 1993.}

\lr\nats{M. Natsuume, \jnl \pr, D50, 3949, 1994.}

\lr \nels {W. Nelson, \jnl \pr, D49, 5302, 1994.}

\lr \chs {C. Callan, J. Harvey and A. Strominger, \jnl \np,  B359,
 611,  1991;
in {\it Proceedings of the 1991 Trieste Spring School on String Theory and
quantum Gravity}, ed. J. Harvey et al. (World Scientific, Singapore 1992).}

\lr \khga {R. Khuri, \jnl \np, B387, 315, 1992;  \jnl \pl, B294,
325, 1992; J. Gauntlett, J. Harvey and J. Liu, \jnl \np, B409, 363, 1993.}

\lr \dukh {M. Duff, R.  Khuri,  R. Minasyan and J. Rahmfeld,
\jnl \np, B418, 195, 1994. }

\lr \napp { M. McGuigan, C. Nappi and S. Yost, \jnl \np,   B375, 421,  1992. }
\lr \sus{ P. Howe and G. Papadopoulos, \jnl \np, B289, 264, 1987;
\jnl \np, B381, 360, 1992. }
\lr \berg { M. Berger, {\sl Bull. Soc. Math. France} {\bf 83}  (1955) 279.}

\lr \sss {T. Banks and L. Dixon, \jnl \np, B307, 93, 1988.}
\lr \gaun{J. Gauntlett, talk presented at the conference
``Quantum Aspects of Black Holes",
Santa Barbara, June 1993.}

\lr \galts {D. Galtsov and  O. Kechkin, ``Ehlers-Harrison-type transformations
in dilaton-axion gravity", preprint MSU-DTP-94/2, hep-th/9407155. }

\lr \howe {P. Howe and G. Papadopoulos,  \jnl \pl, B263, 230, 1991; \jnl \cmp,
151, 467, 1993.}

\lr \suhol{
B. Zumino, \jnl \pl, B87, 205, 1979;
L. Alvarez-G\'aume and D. Freedman, \jnl \cmp, 80, 443, 1981;
S. Gates, C. Hull and M. Ro\v cek, \jnl \np, B248, 157, 1984;
 P. Howe and G. Sierra, \jnl \pl,  B148, 451, 1984;
C. Hull, \jnl \pl, B178, 357, 1986;\jnl \np,  B267, 266, 1986.}
\lr \sev{
P. Spindel, A. Sevrin, W. Troost and A. Van Proeyen, \jnl \np, B308, 662, 1988;
\jnl \np, B311, 465, 1988; B. De Wit and P. van Nieuwenhuizen , \jnl \np, B312,
58, 1989.}
\lr \hus{ C. Hull and B. Spence, \jnl \np, B345, 493, 1990;
E. Witten, \jnl \np, B371, 191, 1992.}

\lr  \bergr{E. Bergshoeff and M. de Roo, \jnl \np, B328, 439, 1989.}
\lr \cand{P. Candelas, G. Horowitz, A. Strominger and E. Witten , \jnl \np,
B258, 46, 1985.}

\lr \susy  {R. Kallosh, A. Linde, T. Ort\'in, A. Peet and A. Van Proeyen,  \jnl
\pr, D46, 5278, 1992.}

\lr \bonn {G. Bonneau and G. Valent, ``Local heterotic geometry in holomorphic
coordinates", PAR/LPTHE/93-56, hep-th/9401003. }

\lr \nonsem{ I. Antoniadis and N. Obers, \jnl \np, B200, 000, 1994}
\lr \sfts{K. Sfetsos and A. Tseytlin,  ``Four Dimensional Plane Wave String
Solutions
with Coset CFT Description", preprint THU-94/08, hep-th/9404063. }
\lr \napwi{ C. Nappi and E. Witten, \jnl \prl, 71, 3751, 1993.}
\lr \kiri {E. Kiritsis and C. Kounnas, \jnl \pl, B320, 264, 1994. }
\lr \kts { C. Klim\v c\'\i k and A. Tseytlin, \jnl \pl, B323, 305, 1994. }
%%%%%%%%%%%%%%%%%%%%%%%%%%%%%%%%%%%%%%%%%%%%%%%%%%%%%%%%%%
\lr \hulla{C. Hull, \jnl \pl, B167, 51, 1986.}
\lr\huto { C. Hull and P. Townsend, \jnl \pl, B178, 187, 1986;
J. Henty, C. Hull and P. Townsend, \jnl \pl, B185,  73, 1987. }

\lr \ross{D. Ross, \jnl \np, B286, 93, 1987.}

\lr \gros{D. Gross and J. Sloan, \jnl \np, B291, 41, 1987.}

\lr \keto{S. Ketov and O. Soloviev, \jnl \pl, B232, 75, 189.}

\lr \barss {I. Bars, ``Curved spacetime geometry for strings and affine
non-compact algebras", USC-93/HEP-B3, hep-th/9309042.}

\newsec{Introduction }
%%%%%%%%%%%%%%%%%%%%%%%%%%%%%%%%%%%%%%%%%%%%%%%

To address strong field effects in string theory, it is necessary to obtain
exact classical solutions and study their properties. As in other field
theories, symmetries have been used to help find these
solutions. It is easy to show that every Killing vector on spacetime
gives rise to a conserved current
on  the string world sheet. If the antisymmetric tensor field is related to the
spacetime metric in  a certain way, these currents are chiral. The
existence of such chiral currents  turns out to simplify the search for
exact solutions. One example
 is the WZW model which describes string propagation on a group
manifold. This background has a large symmetry group, and all the associated
currents
are chiral. (Since the gauged WZW models can be represented in terms of the
difference between
two WZW models for a group and a subgroup, a similar statement applies there.)
Another example is provided by  the $F$-models discussed in \refs{\horts,\hrt}
which have two null Killing vectors and two associated chiral
currents. In addition to these two examples, the only other known exact
solutions to (bosonic) string theory are the plane waves and their
generalizations \refs{\brink,\guv}, which are characterized by the existence of
a  covariantly constant null Killing vector.

We will show that the $F$-models and generalized plane waves
are both special cases of a larger class of exact solutions which have a
null Killing vector and an associated conserved chiral current.
 Backgrounds of
this type are described by  \sms  which  we will  refer to as  ``chiral null
models".
 We will see
that they include a number of interesting examples.

 The presence of a null chiral  current is associated with an
infinite-dimensional affine symmetry of  the \sm action. This
implies  special properties of the  spacetime fields.
The generalized connection with torsion equal to the antisymmetric field
strength plays an important role since it is the one that appears
in the  classical
string equations of motion. We will see that this connection has
reduced holonomy.
A  certain balance between the metric and the antisymmetric tensor resulting in
 chirality  of the action
is the crucial property of our models which is in the core of their exact
conformal invariance.

There are several levels of describing solutions to string theory. The
string equation is usually expressed in terms of  a power series in $\a'$.
If one keeps only the leading order terms, one obtains an equation analogous
to Einstein's equation and a large number of solutions have been found. The
form of the higher order terms is somewhat ambiguous due to the freedom
of choosing different renormalization schemes (or field redefinitions).
For the plane - wave type solutions and the $F$-models, it has been shown that
there exists a scheme in which  the leading order solution does not
receive $\a'$ corrections, and thus corresponds to an exact solution as well.
We will see that the same is true for the more general chiral null models.

To explore the properties of a given solution, one would like to know
not only that a given background is an exact solution to the field equations,
but also what the string states and interactions are in this background. In
other words, one would like to know the corresponding conformal field theory
explicitly. This is known only for  gauged WZW models. But
some chiral null models can be realized as gauged WZW models
\refs{\klts,\horts} so in these
cases, one has more information about the solution.

Many of the chiral null model backgrounds  have unbroken  space-time
supersymmetry
and some  models  admit  extended world sheet supersymmetry. For example,
the $F$-models in even dimensions always have at least (2,0) world sheet
supersymmetry. However,
our argument that they are exact string solutions is not based on this fact.
We will show that these backgrounds are solutions in the bosonic
as well as the superstring and heterotic string theories.
 What types of solutions belong to this
class?  To begin, all of the plane wave type solutions are included, as well
as all of the $F$-models \horts\ which contain the fundamental string solution
\gibb\ as a special case. In addition, several  generalizations of these
solutions are in this class, including the traveling
waves along the fundamental string \garf.
Although the bosonic string does not have fundamental gauge fields,
effective gauge fields can arise from dimensional reduction. In this way,
we will show that the charged fundamental string solutions \refs{\sen,\wald}
are exact.

Perhaps of
most importance is the fact that four dimensional extremal electrically
charged black holes \refs{\gib,\gim,\gar} can be obtained from the
dimensional reduction of
a chiral null model, and hence are exact. Similarly, we will see
that the generalizations of
the extremal black holes which include  NUT charge and rotation
\refs{\kall,\jons,\galts} are also exact. Finally, the chiral null models
also describe some backgrounds with magnetic (and no electric) fields,
as well as
other solutions  which appear to be new.

If one considers only the leading order string equations, many of these
solutions arise as the extremal limit of a family of solutions with a regular
event horizon. The non-extremal solutions are not of the chiral null form
and are likely to receive  $\a'$ corrections in all renormalization schemes.
Finding the exact analogs of these solutions (which include the
Schwarzschild metric as
a special case) remains an outstanding open problem. The fact that
we only obtain a particular charge to mass ratio
from a chiral null model can be
understood roughly as follows.
To have  chiral currents, one needs a balance between the spacetime
metric and antisymmetric tensor field, which upon dimensional reduction
results in a relation between the charge and the mass.

This paper is organized as follows.
In the next section we introduce the chiral null models, and discuss their
properties as well as  some special cases and examples of solutions.
In Section 3  we describe  a general scheme of Kaluza-Klein type dimensional
reduction working directly at the level of the string  world sheet action.
Unlike the
more traditional approach which uses the leading order terms of the spacetime
effective action,  our approach
applies  to all orders in $\a'$.
Section 4 will be devoted to solutions obtained from the dimensional reduction
of a chiral null model. These include the charged fundamental string,
extremal electric black holes and their generalizations.

Section 5 contains our main result: we prove that for a chiral null model,
the leading order solutions do not receive any $\a'$ corrections
(in a particular scheme).
In Section 6 we extend this argument to the case of superstring and heterotic
string theory. We show that  the $(1,0)$ supersymmetric extensions of our
bosonic models are conformally invariant
without any extra  gauge-field  background.
   We also  discuss the
 world sheet supersymmetry properties of these
models.
Section 7   is devoted to  some concluding remarks.

In Appendix A we summarize the geometrical properties
of the
string backgrounds described by the chiral null model (the generalized
connection with torsion, its  holonomy and
curvature tensor, parallelizable spaces, etc.).
 In Appendix B
  we elaborate on the discussion of $D=3$ models in \horts\ and  show that the
general chiral null model in three dimensions is actually a gauged WZW model.

 %%%%%%%%%%%%%%%%%%%%%%%%%%%%%%%%%%%%%%%%%%%%%%%%%%%%%%%

\newsec{Chiral null models: general properties and examples}
%%%%%%%%%%%%%%%%%%%%%%%%%%%%%%%%%%%%%%%%%%%%%%%%%%%%%%%%%%%
\subsec{Review of previous work}

A  bosonic string in a general `massless' background is described  (in the
conformal gauge)
by the \sm
\eqn\mod{  I= {1\ov \pi \a' } \int d^2 z \ L\ , \ \ \ L=  (G_{MN} + B_{MN})(X)\
\del X^M \bd X^N
+ \a'{\cal R}\p (X)\ ,  }
where $G_{MN}$ is the metric,  $B_{MN}$ is the antisymmetric tensor and $\p$ is
the dilaton \frts\
(${\cal R}$ is related to the world sheet metric $\g$  and its
scalar curvature by  $ {\cal R} \equiv \fourth \sqrt \g R^{(2)}$; $\del$ and
$\bd$ stand for $\del_+$ and $\del_-$ when the world sheet  signature is
Minkowskian).

In \horts, two types of models were studied, which were called the \KM and
the $F$-model. In terms of the
coordinates  $X^M=(u,v,x^i)$,
 the simplest (flat transverse  $x^i$-space) $K$- and $F$-model Lagrangians are
\eqn\mok{L_K= \del u \bd v +  K(x) \ \del u \bd u +
\del x_i \bd x^i  + \a'{\cal R}\p_0\ ,  \ \ \ \p_0=\const\  ,  }
\eqn\mof{  L_F= \ F(x) \ \del u \bd v + \del x_i \bd x^i
+ \a'{\cal R}\p (x) \  .  }
These two models are dual in the sense that applying a spacetime duality
transformation \busch\ with respect to $u$ turns the \KM into the \FM
 with $F= K\inv, \ \p= \p_0 + \ha \ln F $.
The general \KM includes arbitrary $u$ dependence and
describes the standard plane fronted waves. It is conformal to all orders
if it is conformal at leading order, i.e.  $\del^2 K=0$.
There  exists a  special scheme  \horts\ in which a  similar statement is true
for  the $F$-model, i.e.  it is conformal to all orders if
\eqn\eqq{   \del^2 F\inv=0 \ , \ \ \   \p=  \p_0 + \ha  \ln F(x)\ .   }
Perhaps the most important solution in this class is the one describing the
fields outside of a
fundamental string  (FS) \gibb\ which is given by
\eqn\fund{
F\inv ={ 1 + {M \ov r^{D-4}} }   \ , \ \ D>4 \ ;\ \  \ \ \
 F\inv ={ 1 - {M \  \ln \ {r\ov r_0} } } \ , \ \ \  D=4\ , }
where $r^2 = x_i x^i$ and $D$ is the total number of space-time dimensions.

The key property of the \KM is that it has a covariantly constant null
vector $\del/\del v $. The main features of the \FM are that there are two
null Killing vectors corresponding to translations of $u$ and $v$, and that
the
coupling to $u,v$ is chiral (since $G_{uv} = B_{uv}$). This means that the
\FM is invariant under the infinite dimensional symmetry $u' =u + f(\tau -
\s)$ and $v' = v+ h(\tau +\s)$. Associated with this symmetry are  two
conserved  world sheet chiral currents:
 ${\bar J}_u = F\bd v, \ J_v = F \del u$.
These properties are preserved if the transverse $x^i$-space  is modified.
In fact,
the  two models \mok\ and \mof\ can be generalized \horts\ to the case
when the transverse space  corresponds to
an arbitrary conformal $\s$-model.  The simplest generalization
is to  keep the transverse metric flat but include an extra  linear
term in the  dilaton.

%%%%%%%%%%%%%%%%%%%%%%%%%%%%%%%%%%%%%%%%%%%
\subsec{The general  chiral null model }
%%%%%%%%%%%%%%%%%%%%%%%%%%%%%%%%%%%%%%%%%%%%%%%%%%%

The fact that a  leading order solution turns  out to be exact applies to
a larger class of backgrounds than represented by
 the \KM and  $F$-model.
 We will consider
the following Lagrangian which will be called the chiral null model:
\eqn\mox{ L=F(x) \del u \bd v   +  \K(x,u)  \del u \bd u +  2\tilde A_i (x,u)
\del u  \bd x^i +  \del x_i \bd x^i     + \a'{\cal R}\p (x,u)\ .  }
We need to assume that $F$ does not depend on $u$ since otherwise
the argument for conformal invariance given in Section 5 does not
go through.
As in the case of the \KM and $F$-model, it is possible to replace the flat
transverse space by an arbitrary conformal $\s$-model, but we will not consider
that generalization here.

This model has roughly half the symmetries of the $F$-model. There is one
 null Killing vector generating shifts of $v$, and
the  action is invariant
under the affine symmetry $v'= v+ h(\tau + \s)$
which is related to the existence of  the
conserved chiral current $J_v= F(x) \del u$.
This in turn implies the special geometrical (holonomy) properties of the
corresponding string backgrounds (see Appendix A).
 Like the $F$-term, the  vector coupling has a special chiral structure: the
$G_{ui}$  and $B_{ui}$ components of the
metric and the antisymmetric tensor are equal to each other.

The action \mox\ can be represented in the form
\eqn\moxa{ L=F(x) \del u \[\bd v   +  K(x,u)   \bd u +  2A_i (x,u)  \bd x^i \]
+  \del x_i \bd x^i     + \a'{\cal R}\p (x,u)\ , }  $$
  K\equiv F\inv \K \ , \ \ \ \ \ A_i \equiv F\inv \A_i \ ,$$
and thus
is invariant under the  subgroup of coordinate transformations $v'=v-
2\eta(x,u)$ combined with  a `gauge transformation'
\eqn\ssa{ K' = K + 2\del_u \eta  \ , \ \ \ \ A'_i= A_i  + \del_i \eta  \ .}
It is clear that using this freedom  one can  always choose a gauge in which
$K=0$. However, we
will often consider the special case when $K,\ A_i$ and $\p$  do not depend on
$u$, i.e. when  $\del / \del u$ is a Killing vector.
In this case, $K$
%is the norm of the Killing vector $\del / \del u$ and
cannot be set to zero without loss of generality.

When the fields do not depend on $u$, one can  perform a
leading-order duality  transformation
along any non-null direction in the $(u,v)$-plane. Setting $v = \hat v + a u$
($a$=const) in \moxa\ and dualizing with respect to $u$ yields a $\s$-model
of exactly the same form with $F,\ K,\ A_i$ and $\p$ replaced by
\eqn\dudu{
 F'=  (K + a)^{-1}\ , \ \ \  K'=  F\inv
 \ , \ \ \  A_i'= {A_i} \ , \
\ \ \p'= \p - \ha \ln [F(K+ a) ] \  \ . }
In other words, chiral null models are `self-dual':
the null translational symmetry and chiral couplings are
preserved under duality.

In Section 5
 we shall determine  the conditions on the functions $F,\ K,\ A_i$ and $\p$
 under which
these models are conformal to all orders in $\a'$. As in the case of the
simplest \FM\ \mof\ there exists a scheme in which these  conditions  turn out
to be equivalent
to the  leading-order equations (derived in Appendix A)\eqn\qwqw{   -\ha \del^2
F\inv +  b^i \del_i F\inv =0 \ , \ \ \ \  -\ha  \del_i {\cal F}^{ij} + b_i
{\cal F}^{ij} =0 \ , }
\eqn\uuu{
 - \ha  \del^2 K  +  b^i \del_i K +    \del^i\del_u{  A_i} - 2b^i \del_u A_i
+ 2 F^{-1} {\del^2_u \p}  =0\ , }
\eqn\qwqww{ \p(u,x)=  \p(u) + b_i x^i + \ha  \ln F(x)\ , }
where
$$  {\cal F}_{ij}\equiv \del_i A_j - \del_j A_i\ , \ \ \ \ \del^2\equiv
\del^i\del_i \ . $$
 Notice that the leading order equations allow  a linear term $ b_i x^i$
in the dilaton.
Eq. \qwqww\ implies that the central charge of the model
is given by $c= D  + 6 b^ib_i$. One can easily verify that these equations
are invariant under the `gauge' transformations \ssa\ (and, when the fields do
not depend on $u$, under  the
duality transformations \dudu). When $F,K,A_i,\p$ are  independent of $u$ and
 $b_i=0$, these equations take the simple form
\eqn\eqqq{   \del^2 F\inv=0 \ , \ \ \  \del^2 K =0 \ , \ \ \   \del_i {\cal
F}^{ij} = 0 \ , \ \ \ \  \p=  \p_0 + \ha  \ln F(x) \ .  }
A crucial feature of these equations is that they are linear. Thus all
solutions
satisfy a solitonic no-force condition and can be superposed (this is also
true for  the more general equations \qwqw\ - \qwqww\ provided $b_i$ is held
fixed).
Since these equations  are  exact conformal invariance  conditions,  changing
$F$, $K$ or $A_i$ while preserving \qwqw--\qwqww\  can be viewed as `marginal
deformations' of the
corresponding conformal field theory.

%%%%%%%%%%%%%%%%%%%%%%%%%%%%%%%%%%%
\subsec{Some special cases}
%%%%%%%%%%%%%%%%%%%%%%%%%%%%%

We now discuss some special cases of the general chiral null model \mox.
If $F=1$, we obtain a class of  plane fronted wave backgrounds which
have a covariantly constant null vector. The general background with
a covariantly constant null vector contains
another vector coupling \brink
\eqn\mokas{  L= \del u \bd v  +   K(x,u) \del u \bd u  +   2A_i (x,u) \del u
\bd x^i  +
2\bar A_i (x,u) \del x^i  \bd u } $$
+  \  \del x_i \bd x^i + \a'{\cal R}\p (x,u)  \ .  $$
The conditions of conformal invariance of this model turn out  to
take the form \tsnul\ (for simplicity we set $b_i=0$)
\eqn\cone{ \del_i {\cal F}^{ij} = 0\ , \ \ \ \  \del_i \bar {\cal F}^{ij} =0 \
,  \ \ \ \p= \p(u) \ , }
\eqn\coni{- \ha  \del^2 K
+   \del^i\del_u( A_i + \bar A_i) +  {\cal F}^{ij} \bar {\cal F}_{ij}
+ 2 \del^2_u\p \  + \   O\(\a'^{s+k} \del^s  {\cal F} \del^k \bar  {\cal F}\)
=0 \ .  }
Thus, if one breaks the chiral structure by introducing the $\bar
A_i$-coupling,   then, in general,
there are corrections  to the $uu$-component of the metric conformal anomaly
coefficient \coni\ to all orders in $\a'$.
 The higher-loop corrections  still  vanish  in one special case: when
$A_i$ and $ \bar A_i$  have   field strengths  constant in $x$ (in general,
the field strengths may still depend on $u$)
\eqn\cobf{ A_i = -\ha  {\cal F}_{ij}  x^j \ , \ \ \ \ \
\bar A_i = -\ha  \bar {\cal F}_{ij}  x^j \ . }
 Such a model  represents a simple and  interesting  conformal theory  in
its own right.\foot{One particular case corresponds to the    $D=4$
non-semisimple WZW
model of ref. \napwi, namely,
 $K= - x^i x_i , \  A_i  = - \bar A_i = -\ha
\ep_{ij} x^j , \  \p=\const$, which is obviously  a  solution of
\cone,\coni.
Since $\bar A_i = -A_i$,   $A_i$ represents  the antisymmetric tensor part
of the action  \mokas. Another equivalent (related by a $u$-dependent
coordinate transformation of $x^i$)  representation
of the model of \napwi\ is $K=0, \ A_i = - \ha \ep_{ij} x^j, \ \bar A_i =0 $
which will be useful
at the end of Section 4 (see also Appendix A).
}
When the fields do not depend on $u$ one may define the dual \sm
which  is also conformal to all orders and will be discussed  at the end of
Section 5.

The special
property of the  model  with $\bar A_i=0$ or $A_i=0$ (i.e. with $G_{ui}=\pm
B_{ui} $) resulting in cancellation of the   vector-dependent contributions to
the
$\beta$-function for $K$  was  noted  at the one-loop
level in \tsnul\ and extended to the two-loop level in \duval.\foot{
It was  observed  in \duval\ that  introducing the generalized connection with
the antisymmetric tensor field strength as torsion,
one finds  that if $\bar A_i=0$
 the generalized curvature (see Appendix A) is nearly flat: the only
non-trivial components of it
are  $\  { \hat R}^v_{- ijk}=2\del_i {\cal F}_{jk}, \
{\hat R}^v_{- iuj}= 2 \del_i\del_u A_j -\del_i\del_j K $.
Then assuming that all terms in the $\b_{\m\n}$-function have the structure
$Y_\m^{\l\r\s} {\hat R}_{- \l\r\s\n}$,   where $Y$ depends on $H_{\m\n\l} $ and
$R_{\m\n\l\r}$
(in a special renormalization scheme this is true  at the 2-loop order \metts)
%and
%should be
% true at higher loop orders as well,  in agreement with the expectation
%%\mukhi\ that
%$ \b_{\m\n}=0$  when ${\hat R}^\l_{\ \m\n\r} =0$)
  one can  argue \duval\ that  all higher-order corrections vanish.
This argument is not completely rigorous and, in fact,    unnecessary,
 since  a  simpler
 direct proof of conformal invariance of  this model can be given  (see Section
 5).}
  It was further  shown  \bergsh\ that such backgrounds
are (`half') supersymmetric  when embedded in $D=10$  supergravity  theory and
it was conjectured   that these  `supersymmetric string waves'
  remain  exact
heterotic  string solutions
to all orders in $\a'$ when supplemented  with some   gauge field background.
  As we  shall demonstrate, \mokas\ with $\bar A_i =0$
is,  in fact,  an exact solution
of the bosonic string  theory.
In Section 6 we shall prove that, furthermore,  it can be promoted to an
exact superstring and heterotic string solution with no need to introduce an
extra  gauge field background.
   It is
the  chiral structure of this solution
  which is behind this fact.

If $K=0$, and $\tilde A_i (x), \ \p$ are independent of $u$,
the chiral null model \mox\ reduces to
\eqn\mofas{  L= F(x) \del u \bd v  +    2 \tilde A_i (x) \del u  \bd x^i
+  \del x_i \bd x^i + \a'{\cal R}\p (x) \ .   }
This background is also
supersymmetric \kallosh\ when embedded in  $D=10$  supergravity theory
(and was also conjectured \kallosh\ to correspond to an exact heterotic
string solution when suplemented by a gauge field).
As above, we will prove in Section 6 that it is  an exact  solution
of the heterotic string theory by itself, i.e.
that the $(1,0)$ supersymmetric extension
of \mofas\ is a conformally invariant model without extra
gauge field terms added.
%Under certain
%conjectures about the structure of $\a'$-corrections it  was suggested in
%\kallosh\ that this background (supplemented by a  Yang-Mills background)  is
%%%a
%solution of the heterotic string theory to all orders in $\a'$.

%%%%%%%%%%%%%%%%%%%%%%%%%%%%%%%%%%%
\subsec{Examples of solutions}
%%%%%%%%%%%%%%%%%%%%%%%%%%%%%

We now discuss some examples of solutions which are described by
chiral null models.
These solutions can be viewed as different
generalizations of the fundamental string solution \fund.

It is straightforward to describe the general solution for the conformal
 $D=5$
chiral null model which is independent of $u$ (and has $b_i=0$). It
is given by
\eqn\fivg{ L=F(x) \del u \[\bd v   +  K(x)   \bd u +  2A_i (x)  \bd x^i \]
+  \del x_i \bd x^i     + \a'{\cal R}\p (x)\ ,  }
where the functions $F,\ K,\ A_i$ and $\p$ satisfy \eqqq. Since the
transverse space is now three dimensional, every solution $A_i$ to Maxwell's
equation  can be written in terms of a scalar\foot{Another simple case is $D=4$
since in two transverse dimensions $A_i = q \epsilon_{ij} x^j$. }
\eqn\trt{  \ep^{ijk}\del_j A_k= \del^i  T (x)\ , \ \ \ \ \   \del^2 T =0 \ . }
With $F\inv$ and $K$ also satisfying   Laplace's equation in the transverse
space,  the general solution is characterized by three harmonic
functions. It is clear from \uuu\ that  the model remains conformal
if  we let $K$  have an  arbitrary
$u$ dependence.
If we set $A_i=0$, take  $F$ and $\p$  given by the FS solution \fund, and keep
$K$ general, the solutions describe
traveling waves along the fundamental string and were first discussed in \garf.

Consider now spherically symmetric solutions with $A_i=0$ and no $u$
dependence. Since all
spherically symmetric solutions to Laplace's equation take the form
$a + br^{4-D}$, the function $K$ can always be represented as $K(x)=c + n F\inv
(x)$.
After a shift of $v$  the model then takes the form \mox\ with $\K=n$. In view
of the freedom to rescale $u$ and $v$ the only non-trivial values of the
constant $n$ are 0 and 1.
$n=0$ corresponds the standard FS  while $n=1$ yields the following simple
generalization
\eqn\mofke{ L=F(x) \del u \bd v +    \del u \bd u  +  \del x_i \bd x^i     +
\a'{\cal R}\p (x)\ ,  }
where $F$ and $\p$ are given by \fund\ and \eqq.
This solution was first found in \wald\
and further discussed in \ghrw.

 It is
known \horstr\  that the fundamental string is the extremal limit
of a family of charged black string solutions to the leading order equations.
The generalization \mofke\
can similarly be viewed as the extremal limit of a black string as follows
(we consider $D=5$ for simplicity). The charged black string  can be
obtained by boosting the direct product of the
Schwarzschild background with a  line,
and applying a duality
transformation \hhs. The result is
($S \equiv \sinh \a, \ C\equiv \cosh \a, $  $\  \a$ is the original boost
parameter)
\eqn\bksth{
 ds^2 = \(1+ {2mS^2\ov r}\)^{-1} \[-\(1-{2m\ov r}\) dt^2 + dy^2\]
       + \(1-{2m\ov r}\)^{-1} dr^2 + r^2 d\Omega \ , }
       $$  B_{yt} = { C\ov S}  \(1 + {2mS^2\ov r}\)\inv
	 \  ,\qquad e^{-2 \phi} = 1+{2m S^2\over r} \ .  $$
 The extremal limit
corresponds to  sending $m\ra 0, \ \a \ra \infty$ in such  a way that $ M\equiv
	  2me^{2\a}$
is held fixed. In this limit the horizon at $r=2m$ shrinks down to zero
size and becomes singular. The charged black string solution \bksth\ approaches
the fundamental string \mof. If we add linear momentum to \bksth\ by applying
a boost
$t=\hat t \cosh \b + \hat y \sinh \b, \ \ y=\hat t \sinh \b + \hat y \cosh \b$,
and then take the extremal limit
$m\ra 0, \ \a,\b \ra \infty$ with
$ M\equiv 2me^{2\a}=2me^{2\b}$ fixed, we obtain the generalized fundamental
string \mofke. So this solution can also be viewed as the extremal limit
of a charged black string, but now with a non-zero  linear momentum.

%%%%%%%%%%%%%%%%%%%%%%%%%%%%%%%%%%%%%%%%%%%%%
\newsec{Dimensional reduction}
%%%%%%%%%%%%%%%%%%%%%%%%%%%%%%%%%%%%%%%%%%%%%%%%%%%%%%%%%%%

To consider  further applications of the
chiral  null models to,  for example,
extremal dilatonic black holes in $D=4$
and charged FS solutions,
we need to  discuss  first the Kaluza-Klein re-interpretation
of higher dimensional bosonic string solutions (heterotic string solutions will
be discussed in Section 6).
To have extremal black holes we need gauge fields. There are no fundamental
gauge fields in bosonic string theory but they appear once the theory is
compactified on
a torus or a group manifold and is expressed in terms of `lower-dimensional'
geometrical
objects.

The usual treatment of dimensional reduction in field theory starts with a
spacetime action.
This is possible also in string  theory, but difficult to do exactly. One would
have to start with the full massless
string effective action  in, say, five dimensions containing terms of
all orders in  $\a'$.
Assuming the fifth direction $x^5$ is periodic we can  expand the metric,
antisymmetric tensor and dilaton  in Fourier  series in $x^5$ and explicitly
integrate over $x^5$.
The result will be the effective action in $D=4$ containing massless fields
as well as  an  infinite tower
of massive modes with masses proportional to a  compactification scale.
Any exact solution of the $D=5$ theory which does not  depend on $x^5$ can then
be
directly interpreted as a solution of the equations of the
$D=4$ `compactified' theory
with all massive modes set equal to zero (but all  `massless' $\a'$-terms
included).

Fortunately, in string theory there is a simpler alternative -- to
 perform the dimensional reduction
directly at the  more fundamental level
of the string action itself.  Let us start with the
general string \sm \mod,
split the coordinates $X^M$  into `external' $x^\m$ and `internal'
$y^a$  and assume that the couplings do not depend on $y^a$,
\eqn\kak{ L= (G_{\m\n} + B_{\m\n}) (x) \del x^\m \bd x^\n +  (\cA_{\m a} +
\cB_{\m a} ) (x)\del x^\m\bd y^a
+ (\cA_{\m a} -  \cB_{\m a} ) (x)\bd x^\m \del y^a }
 $$  + \ (G_{ab} + B_{ab})(x) \del y^a \bd y^b
  +   \a'{\cal R}\p (x) \ , $$
where
 \eqn\defi{ \cA_{\m a} \equiv G_{\m a}\ , \ \
\ \ \ \  \cB_{\m a }\equiv {B}_{\m a  } \ . }
Assuming for simplicity that $B_{ab}=0$, it is easy to represent
the action in a form  which  is manifestly invariant under
the space-time gauge transformations of the vector fields
$\cA^{a}_{\m}\equiv G^{ab} \cA_{\m b} $ and $ \cB_{\m a  }$
\eqn\kaki{ L= \  (\hat G_{\m\n} + B_{\m\n}) (x) \del x^\m \bd x^\n \  +  \
  \cB_{\m a  } (x) (\del x^\m \bd y^a- \bd x^\m \del y^a  )  }
$$ +\  G_{ab}(x)  \[\del y^a + \cA^a_\m (x) \del x^\m\]\[\bd y^b + \cA^b_\n (x)
\bd x^\n\]
 +   \a'{\cal R}\p (x) \ , $$
where  the gauge-invariant  `Kaluza-Klein'  metric  $\hat G_{\m\n}$
 is  defined by
 \eqn\metr{\hat G_{\m\n} \equiv  G_{\m\n} - G_{ab}\cA^a_\m \cA^b_\n
\ . }
Like all $\s$-model Lagrangians, \kaki\ changes by a total derivative
if one adds the curl of a vector to the antisymmetric tensor field. Since
we are assuming no dependence on $y^a$, the $(\m, a)$-component of this
transformation is simply ${\cal B}_{\m a  } \ra {\cal B}_{ \m a  } + \del_\m
\l_a$,
i.e. the standard gauge transformation for the vector fields ${\cal B}_{\m a}$.
The action \kaki\ is also invariant under  shifting $y^a \ra y^a-\eta^a(x)$
together with
\eqn\gaage{\cA^{a}_{\m} \ra \cA^{a}_{\m}+ \del_\m \eta^a\  , \ \ \ \  \
        B_{\m\n} \ra B_{\m\n}-  2 \del_{[\m} \eta^a\cB_{\n] a} \ . }
The first transformation is the usual one for the vector fields $\cA^{a}_{\m}$
while the second
implies that the gauge-invariant
antisymmetric tensor field strength
is given by\foot{From the  world sheet point of view we are using
there seems to be no reason to redefine the antisymmetric tensor  $B_{\m\n}$
in \kaki\
by the term $\cA^a_{[\m}  \cB_{\n]a}$ as it is sometimes done in the effective
action approach  to dimensional reduction.
If one does such a redefinition,  the new  $\hat B_{\m\n}$ also
transforms under the $\cB_{\m a}$ gauge transformations and the generalized
field strength tensor $\hat H_{\l\m\n}$  takes a more `symmetric'  form with
respect to the two vector fields  $\cA^a_\m$ and $\cB_{\m a}$.
It should be noted, however, that it is the full $\hat H_{\l\m\n}$ that has
an invariant meaning, and  it  remains the same irrespective of the definition
of $\hat B_{\m\n}$. }
\eqn\hmnk{ \hat H_{\l\m\n} = 3\del_{[\l} B_{\m\n]} - 3 \cA^a_{[\l} \cB_{\m\n]
a}
\ , \ \ \ \ {\cal B}_{\m\n a} \equiv  2 \del_{[\m} \cB_{\n]a}
\ . }

Although the world sheet approach to dimensional reduction in string theory is
the most straightforward and simplest, it is
 useful to recall what the corresponding procedure  looks like from the
 point of view of the space-time effective action.
For example, if we start with just the
leading-order term in the $D=5$ bosonic string action
\eqn\act{ S_5 = \k_0  \int d^5 x \sqrt G \  \e{-2\p}   \ \{
   \   R \ + 4 (\del_M \p )^2 - {1\ov 12} (H_{MNK})^2\
  + O(\a')   \}  \  , }
  and assume that all the fields are independent of $x^5$,  we obtain the four
  dimensional reduced action
 (for the general case, see  e.g. \maha\ and refs. there)
\eqn\acttp{  S_4 = \hat \k_0\int d^4 x \sqrt {\hat G }\  \e{-2\p + \s}    \ \{
  \   \hat R \ + 4 (\del_\m \p )^2 - 4 \del_\m \p \del^\m \s }
$$  - {1\ov 12} (\hat H_{\m\n\l})^2\  - \fourth \e{2\s} ({ \cal F}_{\m\n})^2
-\fourth  \e{- 2\s} (\cB_{\m\n})^2
  + O(\a')   \}  \  , $$
where  we have defined
\eqn\fgfg{ G_{55}\equiv  \e{2\s}   ,   \ \  {\cal F}_{\m\n} = 2\del_{[\m}
\cA_{\n]}  \ ,   \ \ \cB_{\m\n} = 2 \del_{[\m} \cB_{\n]}  \  , \ \ \ \cA_\m
\equiv    \cA^5_\m\ ,  \ \cB_\m \equiv    \cB_{\m 5}\  . }
Setting \eqn\iii{\vp = 2\p -  \s\     }
the action \acttp\ becomes
\eqn\acttq{  S_4 = \hat \k_0\int d^4 x \sqrt {\hat G }\  \e{- \vp}    \ \{
  \   \hat R \ +  (\del_\m  \vp )^2 - (\del_\m \s)^2  }
$$  - {1\ov 12} (\hat H_{\m\n\l})^2\  - \fourth \e{2\s} ({\cal F}_{\m\n})^2
-\fourth  \e{- 2\s} (\cB_{\m\n})^2
  + O(\a')   \}  \  .  $$
In
 the Einstein frame \acttq\ takes the form
\eqn\acttw{S_4 = \hat \k_0\int d^4 x \sqrt {{\hat G_E}}   \ \{
  \   \hat R_E \  - \ha  (\del_\m  \vp)^2 - (\del_\m \s)^2 } $$
 - \ {1\ov 12} \e{ - 2\vp  }   (\hat H_{\m\n\l})^2\
- \fourth \e{- \vp  +  2\s } ({\cal F}_{\m\n})^2
-\fourth  \e{- \vp -  2\s}  (\cB_{\m\n})^2
  + O(\a')   \}  \  . $$
Thus, in general, the four dimensional theory contains two scalars, two
vectors,
and the antisymmetric tensor, as well as the metric.
In certain special
cases, the nontrivial part of the action \acttw\ can be expressed in terms of
only one scalar and one vector,
so that it takes  the familiar form\foot{Such ansatzes must,
of course,  be consistent with $D=5$ equations of motion.}
\eqn\actt{S_4 = \hat \k_0\int d^4 x \sqrt {{\hat G_E}}   \ \{
  \   \hat R_E \ - \ha (\del_\m \psi )^2 - \fourth \e{- a\psi} ({\cal
F}_{\m\n})^2  + O(\a') \}  \  . }
For example, if one sets $\p=0$  and $H_{MNK} = 0$ in the $D=5$ action,  or
equivalently $\vp=-\s, \ {\hat H}_{\m\n\l} =0=\cB_{\m\n}$  directly in
\acttw, one
obtains \actt\ with $\psi= - a  \s$ and $ a=\sqrt 3$. This is, of course,
the standard Kaluza-Klein reduction of the Einstein action. Another possibility
is to set $\s =0\
(G_{55}=1),\ \ \hat H_{\m\n\l} =0$ and either the two vector fields
proportional
to each other, or let one of them vanish. This case corresponds to \actt\
 with $\psi= \vp$  and $a =1$.

%%%%%%%%%%%%%%%%%%%%%%%%%%%%%%%%%%%%%%%%%%%
\newsec{Solutions involving dimensional reduction }
%%%%%%%%%%%%%%%%%%%%%%%%%%%%%%%%%%%%%%%%%%%%
In this section we   discuss
the  dimensional reduction of some of the exact
 solutions described by chiral null models \mox. We will see that
 several previously found solutions of the leading order string effective
equations
 can be easily obtained in this way. In addition,  we find some solutions which
 appear to be new.

%%%%%%%%%%%%%%%%%%%%%%%%%%%%%%%%%%%%%
\subsec{Charged fundamental string  solutions }
%%%%%%%%%%%%%%%%%%%%%%%%%%%%%%%%%%%%%%%%%%%%%%%%%
Our first example is  the charged FS solution found at the leading order
level in \refs{\sen,\wald}.\foot{The method of \sen\
was  to start with the neutral solution and to make the
most general leading order duality rotation
in all  available isometric directions (including the internal ones).
Since the duality transformation has, in general,  $\a'$-corrections,
this  procedure does not guarantee the exactness of the resulting solution.
}
This solution is obtained by starting with the general chiral null
model in $D+N$ dimensions, and requiring that all fields be independent
of $u$ and $N$ of the transverse dimensions labeled by $y^a$. If we further
assume that the vector coupling has only $y^a$-components, we obtain
\eqn\mofkac{ L=F(x) \del u \bd v +  \K(x)  \del u \bd u  +
\del x_i \bd x^i     +  2 \tilde A_{a } (x) \del u \bd y^a    + \del y_a \bd
y^a +   \a'{\cal R}\p (x)\ ,   }
which is conformal to all orders provided  $F,\ K\equiv F\inv \K,\
 A_{a}\equiv F\inv \tilde A_{a}$ and $\p$ satisfy
\eqqq.  If we are looking for  FS-type solutions which are rotationally
symmetric
in $D-2$ coordinates $x^i$, then  solving  the Laplace equations  we can put
the functions $F,K, A_a$ in the form\foot{In the zero charge $Q_a=0$ limit
we get not just the  FS solution of \gibb\ but its modification \mofke\
which corresponds to   momentum
running  along the string.}
 $$ F\inv    = 1 + { M \ov r^{D-4}}\ , \ \ \ \
\p  = \p_0 + \ha \ln F(r) \ , \ \ \  r^2=x^ix_i\ ,   $$
 \eqn\lapl{  K= c + { P \ov r^{D-4}}\ , \ \ \ \
 A_{a}  = {Q_a \ov r^{D-4}} \ .  }
Shifting $v$  we can thus in general replace $\K$ in \mofkac\ by a constant.
To re-interpret \mofkac\ as a $D$-dimensional model coupled to $N$ internal
coordinates we rewrite it in the form \kaki\
\eqn\mofkacc{ L=F(r) \del u \bd v +   \K' (r)  \del u \bd u  +
\del x_i \bd x^i +   \a'{\cal R}\p (r) }
 $$  + \   \tilde A_{a} (r)( \del u \bd y^a  - \del y^a \bd u)    + \  [\del
y^a  + \tilde A^{a}(r) \del u ][\bd y_a + \tilde A_{a}(r) \bd u]   \ , $$
$$  \K'(r) \equiv \K - (\tilde A_{a})^2 \ .  $$
The first four terms give the $D$-dimensional space-time metric, antisymmetric
tensor and dilaton  while the last two identify (see \kaki)  the presence of
two equal vector field backgrounds
(two equal  components  $G_{ua}$ and $B_{ua}$
 conspire as one $D$-dimensional
Kaluza-Klein vector field, cf. \acttq).  Note that since $G_{ab} =
\delta_{ab}$, the modulus field is constant and the lower
dimensional dilaton is the same as the higher dimensional one.
%Redefining $v$  we can put $\K'$ into the form $\K'=d_1 + d_2 F^2 , \ d_1=
%%%PM\inv -Q^2_a M^{-2} , \ d_2= - Q^2_a M^{-2}$.
In the case of just one internal dimension we get
 one  abelian  vector field  $u$-component
and the
 resulting background becomes
that of the  charged FS  in \refs{\sen,\wald}.

%%%%%%%%%%%%%%%%%%%%%%%%%%%%%%%%%%%%%%%%%%%%%%%%%%%%%%%
\subsec{$D=4$ solutions with electromagnetic fields}
%%%%%%%%%%%%%%%%%%%%%%%%%%%%%%%%%%%%%%%%%%%%%%%%%%%%%%%%%

To obtain four dimensional solutions with electromagnetic fields,
we can reduce a $D=5$ chiral null model. It was recently shown \hrt\
that extremal electrically charged black holes can be obtained in this way.
If one starts with the standard $D=5$ FS \mof,\fund\   one gets \gaun\ the
extremal electric black hole solution to \actt\ with
$a=\sqrt 3$ which was discussed in
\gib, while starting with the generalized FS  \mofke\
one obtains the extremal electric
black hole solution to \actt\ with $a=1$
discussed in  \refs{\gim,\gar}.\foot{The $a=\sqrt 3$ black
hole can also be obtained \gib\
from the $D=5$ plane-wave-type background \mok\  which is dual to  FS.
Similarly,  one can get the  $a=1$ electric dilatonic  $D=4$
black hole  from  a duality-rotated \dudu\
version of the generalized FS  \mofke.  Such model is,  however,
essentially equivalent to \mofke, since  it is
 `self-dual'.}

 Here we shall   consider   the most general $D=5$ chiral null model
 which is independent of $u$. It
 will yield a  large  class of
$D=4$ solutions. Some of these  backgrounds were recently found
\refs{\kall,\jons,\galts}
as
leading-order  string solutions, i.e. solutions   of the
dilaton-axion  generalization of the $D=4$  Einstein-Maxwell theory.
They   are   the analogs of the   IWP (Israel-Wilson-Perj\'es \iwp) solution of
the pure Einstein-Maxwell theory.\foot{
 It was shown  also that these backgrounds are supersymmetric
when embedded in a supergravity \kall.}
Special cases of this generalized  IWP  solution describe
a collection of extremal electric dilatonic black holes
(Majumdar-Papapetrou-type  solution)  and an extremal electric Taub-NUT-type
solution.

The $D=5$
chiral null model which is independent of $u$
\eqn\yyy{ L_5=F(x) \del u \[\bd v   +  K(x)   \bd u +  2A_i (x)  \bd x^i \]
+  \del x_i \bd x^i     + \a'{\cal R}\p (x)\   , }
was discussed in section 2.4 where it was noted that the general
solution depends  on the three harmonic functions  $F\inv, \  K$ and $T$ (see
\trt) of the three coordinates $x^i$. This model can be reduced to $D=4$
along any space-like direction in the $u,v$ plane.
Shifting $v$ by a multiple
of $u$ changes, of course, the direction of $\del/\del u$,
but this transformation
 is equivalent to a shift of $K$ by a constant.
Shifting $u$ by a multiple of $v$ can be
undone by a particular case of the gauge transformation
\gaage\ (which gives an equivalent   background, in particular,
leaves  $\hat H_{\m\n\l}$ invariant).
Thus  it suffices to use
$u$ as the internal coordinate $y$ (which is possible,  provided
$FK>0$) and to identify $v$ with
$2t$. Then   we  can put   \yyy\ in the
``four-dimensional" form \kaki\ as follows
\eqn\mofee{ L_{5}=- {K}(x)\inv  F (x)
 \[\del t + A_i (x) \del x^i\] \[ \bd t  + A_i (x) \bd x^i\]
+   \del x_i \bd x^i  + \a'{\cal R}\p(x)}
$$   + \ F(x)(\del y \bd t - \del t \bd y)  +
F(x) A_i(x) (\del y \bd x^i - \del x^i \bd y)  $$
$$
  + \  K(x)F(x) \[ \del y +  K\inv (x)   \del t +  K\inv (x)   A_i  (x)  \del
x^i \]$$ $$  \times \[\bd y +
  K\inv  (x) \bd t + K\inv (x)   A_i (x)  \bd x^i \] \
 .    $$
The corresponding  {\it four-dimensional } background is thus  represented by
the following metric,
two abelian
 gauge fields $\cA^5_\m\equiv \cA_{\m}, \
\cB_{\m 5}\equiv \cB_{\m}$, two scalars (the  `modulus' $\s=\ha \ln G_{55}$
and  the dilaton) and the antisymmetric tensor field strength $\hat H$  (cf.
\kaki, \acttq)
\eqn\didii{ ds^2 = - {F(x)  K\inv (x)  }  \[ dt +  A_i(x)  dx^i\]^2 + dx_idx^i
\ , } $$
\cA_t=  K\inv (x) \ ,   \ \ \ \cA_i =  K\inv (x)  A_i(x) \ , \ \ \
\ \ \   \cB_t= -
F(x)\ , \ \ \ \cB_i =-   F(x)  A_i(x) \ , $$ $$   \  \s= \ha \ln [F(x)K(x)]\
, \ \ \  \ \ \p = \p_0 + \ha \ln F(x) \  , \ \ \ \ \
\hat H_{\l\m\n} =  - 6 \cA_{[\l} \del_\mu \cB_{\n]}\ . $$
Notice that even though the $D=4$ antisymmetric tensor  $B_{\m\n}$ vanishes,
the gauge invariant field strength $\hat H_{\l\m\n}$ is nonzero due to the
contribution
from the gauge fields in \hmnk. This background represents  a solution  of  the
equations following from the $D=4$ effective action  \acttq\
since
$A_i$ satisfies $ \ep^{ijk}\del_j A_k= \del^i  T (x)$,  and  $F\inv, K$ and $T$
are solutions of the three dimensional
Laplace equation.

Let us now consider some special cases.
If $K=1$ and $A_i=0 $,   the gauge field $\cA_\m$ becomes trivial and
the two scalars  coincide   (up to a constant).  Since the
gauge fields have only time components being  nonzero, the antisymmetric tensor
$\hat H$ vanishes. If we now set $F\inv =1 + M/r$,
the original $D=5$ theory \yyy\ describes the fundamental string
and the $D=4$ reduction is the
`Kaluza-Klein' extremal  black hole,  i.e. the extreme electrically charged
black hole solution corresponding to \actt\ with $a=\sqrt 3$.
We see that this solution has a straightforward generalization to the
case of $A_i\ne 0$.

The case $K=F\inv$
is of particular interest.
The $D=5$  model \yyy\  is the $A_i$-generalization of  \mofke\  while the
corresponding $D=4$
background is
\eqn\dididi{ ds^2 = - {F^2(x)  }  \[ dt +  A_i(x)  dx^i\]^2 + dx_idx^i \ ,
  }
$$\cA_t=  F(x)\ ,  \ \ \cA_i  = F(x)  A_i(x) \ , \ \ \  \ \cB_\m= - \cA_\m\ ,
 $$
$$   \p = \p_0 + \ha \ln F(x)\ , \ \ \  \ \   \hat H_{\l\m\n} =
 6 \cA_{[\l}
\del_\mu \cA_{\n]}  \ , \ \ \ \ \  \s=0\ .   $$
 Since $\s=0$ and the two gauge fields differ only by a sign, these backgrounds
 are solutions to \actt\ with $a=1$ provided  the antisymmetric tensor  term
of \acttw\ is included.
These are precisely the $D=4$  dilatonic IWP solutions
\refs{\kall,\jons,\galts}.
 If we restrict further to $A_i=0$ and $F\inv = 1 + M/r$, then $\hat
H_{\l\m\n}=
 0$ and
we obtain the `standard' extremal  dilatonic black hole \refs{\gim,\gar}\foot{
Let us note that the $D=4$ extremal electric dilatonic black hole background
can also be related to a $D=6\ $ chiral null model with $K=0$,
$\ \ L_6 = F(x) \del u \[\bd v + 2A(x)\bd y'\] - \del y' \bd y' +  \del x_i \bd
x^i, \  $
where the internal coordinate $y'$ has the `wrong' (time-like) signature.
Introducing the new coordinate $y' = y + u$
and choosing $A= F\inv$ (which is consistent with the conformal invariance
conditions)
 we find that  this model takes the form of \mofke\ plus an extra free
time-like direction,
$\  L_6 = F(x) \del u \bd v +   \del u \bd u  +  \del x_i \bd x^i   - \del y
\bd y , \  $
and thus  can also be related to the $D=4$ extremal electric black hole.
An equivalent  observation  was  made  at  the level of the leading-order terms
in the effective action in  \bergkal\ (ref. \kall\ also discussed  a similar
higher (six)  dimensional interpretation
 of the IWP solution). It should be emphasized that it is our
 $D=5$ model \yyy\ that provides the correct  higher-dimensional
embedding of these $D=4$ black-hole type solutions:
though the presence of an extra  time-like `internal'  coordinate in the above
$D=6$ model
is irrelevant from the point of view of the proof of exactness of
the $D=4$  solution,   it  is unphysical,  since
complex coordinate transformations
are needed if one wants to keep the physical signature of the full
higher-dimensional space.
  }
\eqn\did{ ds^2 = - F^2 (r) dt^2 + dx_idx^i \ ,  } $$
\cA_t=-\cB_t = F(r)\  , \ \ \ \ \ \p(x) = \p_0 + \ha \ln F (r)\  , \ \ \
  \ \ \cA_i = \cB_i =  \s =  \hat H_{\l\m\n} =  0 \ .  $$
Adding a nonzero $A_i$ to this solution  by setting $T = q/r$ has the effect
of adding a  NUT charge. The result is the extremal electrically
charged dilatonic Taub-NUT solution. Linear superposition of an arbitrary
number of solutions of this type is possible by setting
\eqn\maj{ F\inv = 1+ \sum_{k=1}^{N} {M_k\ov |x - x_k|}\ , \ \ \ \  \ \
T =   \sum_{k=1}^{N} {q_k\ov |x - x_k|}\ . }
To add angular momentum, one takes solutions to Laplace's equation which
are singular on circles, rather than points as in \maj.

 Finally, if we set  $K=F=1$ in \dididi, the dilaton becomes constant. This
 solution  depends only on $A_i$ and
 describes a spacetime with a magnetic field $ {\cal F}_{ij} = 2 \del_{[i}
A_{j]}$ and antisymmetric tensor $ \hat H_{tij} = {\cal F}_{ij}$.
The corresponding $D=5$
exact conformal  \sm \yyy\  can be put (by a shift of $v$)
in the following  simple form
\eqn\yyys{ L = \del u \bd v   + 2 A_i(x) \del u \bd x^i
+  \del x_i \bd x^i   \  ,  \ \ \  \ \del_i{\cal F}^{ij} =0 \ ,   }
and deserves further study.
Some special choices of $A_i$ are particularly interesting.
One is the monopole background, $ {\cal F}_{ij} = q \ep_{ijk} x^k/|x|^3$.
Another is the case of  a uniform  magnetic field,  ${\cal F}_{ij}=\const,  $
i.e.
$A_i= -\ha {\cal F}_{ij} x^j$.
 This   model is equivalent  (see Appendix A.3)
to a
 product of the non-semisimple $D=4$ WZW
model of \napwi\ and an extra free
spatial direction and thus has a CFT interpretation.
 One can choose coordinates so that the $D=4$
metric
for the uniform magnetic field solution is simply
\eqn\rotma{  ds^2 =  -\(dt + {1\ov 2}{\cal H}
  r^2 d\theta\)^2 + dz^2 + dr^2
	     + r^2 d\theta^2  \ ,  }
and describes  a rotating universe (while the antisymmetric tensor $\hat H$
is constant).
 This  uniform magnetic field solution
   may be contrasted with  the dilatonic Melvin
solution \refs{\gim,\dgkt} in which the magnetic field decreases with
transverse
 distance. The latter solution    contains
 a nonconstant dilaton (but no antisymmetric tensor or rotation)
 and is expected to
 have higher order $\a'$ corrections.

The   solutions \didii\ with  generic  $K$
and thus different gauge fields $\cA_i$ and $\cB_i$
appear to be new.

%%%%%%%%%%%%%%%%%%%%%%%%%%%%%%%%%%%%%%%%%%%%%
\newsec{Conformal invariance of  the chiral null models }
%%%%%%%%%%%%%%%%%%%%%%%%%%%%%%%%%%%%%%%%%%%%%
The aim of this section is to  demonstrate that the
general chiral null model \mox\ is conformal to all orders in $\a'$
provided the couplings satisfy the conditions \qwqw\ - \qwqww\ and
one choses a special  renormalization scheme.
Our discussion will  be based on the approach of \horts\
where  more details  about the special  choice of the  scheme can be found.

In \horts\ it was  shown that the \FM  \mof\ (i.e. \mox\ with $K=A_i=0$)
which has two null Killing vectors and two associated
chiral currents,  is  exact.
It turns out that a single chiral current associated with a null symmetry
is, in fact,  sufficient to establish the exact
conformal invariance  of the more general backgrounds  \mox.

To find the conditions for conformal invariance of a \sm  we must define it on
a curved two dimensional surface, introduce sources for the \sm fields
and determine when the resulting generating functional (or its Legendre
transform)
does not depend on the conformal factor of the 2-metric.
There are two  reasons why the models \mox\ are special. First,  the
null symmetry and chiral coupling to $v$ imply that the
path integral over $v$ is readily computable giving a $\d$-function
constraint on $u$ which
expresses $u$  in terms of $x^i$ and a source. Second,
chirality of the $\del u\del x$-coupling implies that the resulting
effective $x$-theory  has only tadpole divergences (or conformal anomalies)
in a properly chosen scheme.

We shall first give the proof of  conformal invariance
 in a few special cases  mentioned in section 2
 (when some of the functions in \mox\ are trivial) and then give the general
 argument.

%%%%%%%%%%%%%%%%%%%%%%%%%%%%%%%%%%%%%%%%%%%%%%%5
\subsec{$F$=1}
%%%%%%%%%%%%%%%%%%%%%%%%%%%%%%%%%%%%%%%%%%%%%

The argument is simplest   when  $F=1$.
 To find the  exact  conditions of conformal invariance we  follow \horts\  by
adding the source terms
($z$ denotes the two world sheet coordinates)
\eqn\sou{ L_{source} =  V(z)\del\bd u + U(z)\del\bd v  +  X_i(z)\del\bd x^i  \
, }
to \mox\ and performing  the path integral over $v$. The resulting
$\d$-function
sets
 $u$ to its classical value $U$ (up to a zero mode which we absorb in $U$).
Thus $u$ is `frozen' and
the  effective $x$-theory  is
\eqn\mokaa{ L_{eff}=\del x_i \bd x^i   +  K(x,U) \del U \bd U  +   2A_i (x,U)
\del U  \bd x^i
+ \a'{\cal R}\p (x,U) } $$   + \  X_i\del\bd x^i  + V\del\bd U\   .  $$
Computing  the classical dilaton contribution   ($\sim \del\bd \p$)
to the trace  of the stress energy tensor  and observing that there
cannot be  $O(\bd U\del x^i)$ quantum contributions
(in view of the  absence of the $O(\bd U)$ vector coupling  and simple
dimensional considerations) one finds  that the
 necessary conditions for this  theory to be  conformal  are $\del_i\del_u
\p=0, \ \del_i\del_j \p=0$,  so that
\eqn\dil{ \p(x,u) = \p (u) + b_i  x^i \ , \ \ \ b_i=\const \ .  }
One also learns
that (in the minimal subtraction scheme)
the renormalization of the  $\del U \bd U $  and $\del U  \bd X^i$
may come only  from the one-loop tadpole diagrams. The conclusion is that  this
model  is conformal
to all orders once the leading-order conditions of conformal invariance
are satisfied
(see also \tsnul)
\eqn\lea{ - \ha  \del^2 K  +  b^i \del_i K +
 \del^i {\del_u  A_i} - 2 b^i \del_u A_i
+ 2 {\del^2_u \p}  =0\ , \ \ \  -\ha  \del_i {\cal F}^{ij} + b_i {\cal F}^{ij}
=0 \  .}
These relations follow from a direct computation of the tadpole graphs
and use of classical \sm equations  to transform the dilaton contribution
(for simplicity, one may gauge away $K$ by using the freedom \ssa).
They agree, of course, with  the standard general expression for the one loop
Weyl anomaly coefficients given in Appendix A.

%%%%%%%%%%%%%%%%%%%%%%%%%%%%%%%%%%%%%%%%%%%%%%%5
\subsec{$A_i=0$}
%%%%%%%%%%%%%%%%%%%%%%%%%%%%%%%%%%%%%%%%%%%%%
Let us now set $A_i=0$ and assume that  $K=F\inv \K$ and $\p$  do not depend on
$u$, i.e. consider
\eqn\mofk{ L=F(x) \del u \bd v +  \K(x)  \del u \bd u  +  \del x_i \bd x^i
   + \a'{\cal R}\p (x)\ .  }
Introducing the source terms  \sou\ and integrating over $v$
one finds the constraint
\eqn\cont{\del u = F\inv (x) \del U\ . }
 Integrating then over
 $u$  and taking into account the determinant contribution that shifts the
dilaton as well as fixing the  same special `leading-order' scheme (related to
the standard one by  an $\a'$-redefinition  of the $ij$-component of the
metric)
as in the $F$-model \horts\  one finds  that   the  effective  $x$-theory
takes the form\foot{Note that if  $F$ were $u$-dependent
the integral over $u$ would not be easily computable
and the argument below would not apply.}
\eqn\mofak{ L_{eff}=\del x_i \bd x^i   -  F\inv(x) \del U \bd V
 + \
  K (x) \del U  \bd  \del^{-1}  [ F\inv (x) \del U ] }
$$ + \  \a'{\cal R}\p' (x)   + \  X_i\del\bd x^i \   ,  $$
\eqn\ggg{ \p'\equiv \p -\ha \ln F (x)  \ . }
The  conditions of exact conformal invariance include the  linearity of the
dilaton $\p'$ in $x$
\eqn\coo{ \p'= \p_0  + b_i x^i\ ,  \ \  \ \ \ \p= \p_0  + b_i x^i + \ha \ln F\
, }
and the  standard scalar  (`tachyonic')  equation for $F\inv$
\eqn\rrr{ -\ha \del^2 F\inv  + b^i\del_i F\inv =0\ .   }
The conformal anomaly must be local, so it is  only the local part of the
quantum  average of the non-local  $O(\del U \del U)$
 term that may contribute to it.
Since this non-local term  already contains two factors of $\del U$
it cannot produce  $\del x$-dependent counterterms. That means  we may
expand the functions $ K (x)$ and $F\inv (x) $ in it near a constant,
 $\ x^i(z)= x^i_0 + \eta^i(z),  $
\eqn\xxx{\int d^2 z d^2z' [  K (x)  \del U](z)  {\bd^2  \Delta\inv (z, z') } [
F\inv (x) \del U ]  (z') } $$
=   \sum_{n,m=0}^{\infty} {1\ov n! m!}  \del_{i_1}...\del_{i_m}   K (x_0)
\del_{j_1}...\del_{j_n} F\inv (x_0) $$ $$
 \int d^2 z d^2 z' (\eta^{i_1} ... \eta^{i_m})(z) \del U (z) {\bd^2
 \Delta\inv (z, z') }  (\eta^{j_1} ... \eta^{j_n})(z') \del U (z') \ , $$
where we defined $\Delta\inv$ by $\del\bd \Delta\inv = \delta^{(2)} (z,z')$.
Then the only contractions of the quantum fields $\eta^i$ that can produce
local
$O(\del U \bd U)$
 divergences  are the one-loop tadpoles on the left and right side of the
non-local
propagator $ \Delta\inv (z, z')$.
Any contraction between $\eta^n (z)$ and $\eta^m (z')$
gives  additional $ \Delta\inv (z, z')$-factor and thus contributes only
to the non-local part of the corresponding $2d$ effective action.

As a result, we find the following conformal invariance condition
\eqn\conn{  F\inv \del^2   K   +  K\del^2 F\inv =2b^i F\inv \del_i K
 + 2b^i K \del_i F\inv \ , }
or,  combined  with \rrr,
\eqn\eqqq{   \del^2 F\inv= 2b^i \del_i F\inv \ , \ \ \  \del^2 K = 2b^i \del_i
K\ , \ \ \   \p=  \p_0 + b_i x^i
+ \ha  \ln F\ .   }

%%%%%%%%%%%%%%%%%%%%%%%%%%%%%%%%%%%%%%%%%%%%%%%5
\subsec{General chiral null model}
%%%%%%%%%%%%%%%%%%%%%%%%%%%%%%%%%%%%%%%%%%%%%

For the general chiral null model (with $u$ dependence), one can set
$K=0$ by  the gauge transformation \ssa.
Adding sources and integrating over $v$ and $u$ as above
we arrive at the following effective $x$-theory
\eqn\mofaa{ L_{eff}=\del x_i \bd x^i   - F\inv(x) \del U \bd V  +   2A_i \( x,
\del\inv [F\inv (x) \del U] \) \del U  \bd x^i  } $$
+ \  \a'{\cal R}\p'\( x, \del\inv [F\inv (x) \del U] \)  + \  X_i\del\bd x^i \
 , $$
where  $\p'$ is as in \ggg\ and we again   use
 a special scheme to keep the free kinetic term of $x^i$ unchanged (see
\horts).
The condition of conformal invariance  in the  $ \del x \bd x$ direction  is
straightforward generalization of \dil\ and the condition in  the model with
$A_i=0$ \coo, i.e.
$\p'= \p(u)  + b_i x^i$. The  $\del U \bd V $ term is conformally invariant,
provided    one imposes \rrr\  as in the $A_i=0$ model.
The conditions of conformal invariance in the $\del u\bd u$ and $\del u \bd x$
directions
 are   similar to    \lea\ with $K=0$,
\eqn\leaa{
  { \del^i \del_u A_i} - 2 b^i \del_u A_i
+ 2 F\inv  {\del^2_u \p}  =0\ , \ \ \ \ \ \    -\ha  \del_i {\cal F}^{ij} + b_i
{\cal F}^{ij} =0 \ . }
 The reason why there are no extra terms involving $F$ is that
the locality of the  conformal anomaly
implies that   the only contributions depending on  derivatives of  $F$
are  tadpole ones which thus vanish due to \rrr.
This is  easy to see by expanding the argument $x^i(z) $  of   $F\inv$ and
$A_i$
near its  `classical' value.  Contractions of the quantum fields on
the opposite sides of the $\del\inv$-operator produce only non-local
contributions to the
corresponding effective action.

Equation  \leaa\ is valid in the gauge $K=0$. The general form of
this conformal invariance condition can be obtained by doing the
gauge transformation \ssa. Combining all the conditions together we
obtain\foot{Let us note that the  fact that the model \moxa\ is Weyl invariant
means also  that  when
considered on a  flat  world sheet  this \sm  is ultra-violet finite
to all loop orders  {\it on the mass shell}. The latter clarification means
that the standard $\b$-functions vanish only modulo a diffeomorphism term
(which is related to  the presence of a non-trivial dilaton in the
corresponding  Weyl-invariant model).}
\eqn\qeq{   -\ha \del^2 F\inv +  b^i \del_i F\inv =0 \ , \ \ \ \ \    \p=
\p(u)  + b_i x^i
+ \ha  \ln F(x)\ ,    }
\eqn\qeqp{ - \ha  \del^2 K  +  b^i \del_i K +    {\del^i  \del_u A_i} - 2b^i
\del_u A_i
+ 2 F^{-1} {\del^2_u \p}  =0\ , \ \ \ \  -\ha  \del_i {\cal F}^{ij} + b_i {\cal
F}^{ij} =0 \  .}

%%%%%%%%%%%%%%%%%%%%%%%%%%%%
\subsec{Further generalizations?}
%%%%%%%%%%%%%%%%%%%%%%%%%%%%%%%%%%%%%%
Can one extend the chiral null model \mox\ to include a larger class
of backgrounds
and maintain their conformal invariance? As we have already remarked, one
possible generalization is to replace the transverse space with a nontrivial
conformal field theory. Another possibility would appear to be the
addition of a second
vector coupling
\eqn\moxs{ L=F(x) \del u \bd v   +  \K(x,u)  \del u \bd u +  2\tilde A_i (x,u)
\del u  \bd x^i  }
$$  + \ 2\tilde  S_i(x,u) \del x^i \bd v +   \del x_i \bd x^i     + \a'{\cal
R}\p (x,u)\ .  $$
This $\s$-model shares with the chiral null model the following three
properties:

(i)  conformal invariance of the transverse  part of the model;

(ii) existence of an affine symmetry $v'=v + h(\tau + \s)$
in a  null direction;

(iii)  chirality of all vector couplings.

The second condition implies the existence of  the associated conserved chiral
current. At the `point-particle' (zero mode) level  this affine stringy
symmetry reduces to  the  null Killing symmetry $v'=v + h, \ h=\const$.

However, the model \moxs\ is not, in general, conformal to all orders if
only the leading-order equations  are satisfied.
As  before, we can still explicitly integrate out $v$ and then $u$.  But the
result is a complicated $x$-theory for which the conditions of conformal
invariance  seem
difficult to  formulate and solve  explicitly to all orders.\foot{We assume
that $\K$ or $\tilde A_i$ do not vanish at the same time. In the special case
when
$\K=0$ and $\tilde A_i=0$ the  model \moxs\ is equivalent to the  special case
\mofas\  of \mox\ with $u\ra v$, $v\ra u$. }

To illustrate this point,  let us consider  a particular example
of \moxs\ with $F=1$,  $ \  \tilde A_i=0$ and $u$-independent couplings,
\eqn\moxsw{ L=\del u \bd v   +  \K(x)  \del u \bd u  +  2 \tilde S_i(x) \del
x^i \bd v +   \del x_i \bd x^i     + \a'{\cal R}\p (x)\ .  }
The  corresponding target space metric has a
null Killing vector, but in contrast to the case
of the model \mox\ with $F=1$ this vector is not covariantly constant.
Making the  coordinate transformation $u\ra  u + p(x)$ we get
\eqn\moxsw{ L= \del u \bd v   +  \K \del u \bd u
+ \K \del_i p (\del u  \bd x^i + \del x^i   \bd u) }
$$ \  +  \
(2\tilde S_i + \del_i p)
\del x^i \bd v +   (\delta_{ij} + \K \del_i p \del_j p) \del x^i \bd x^j     +
\a'{\cal R}\p (x)\ .  $$
If we now choose $\tilde S_i= - \ha \del_i p$, the  new $\del x \bd v$-coupling
disappears.
We learn that in this case  the model \moxsw\ is equivalent
to a modification of \mox\ with a non-trivial transverse metric and
non-chiral  $\del u \bd x$  and $\del x \bd u$ - couplings (cf. \mokas).
Integrating over $v$ it is easy to see that the
the resulting conformal invariance conditions  (both in $\del u \bd u$  and
$\del x \bd x $ directions)  contain non-trivial corrections to all orders in
$\a'$.

This example makes it clear that the above three conditions are not sufficient
to ensure that leading order solutions will be exact.
One needs an additional condition
which can be taken to be

(iv) the null Killing vector should be orthogonal to the transverse subspace.

One can further generalize \moxs\ by introducing a non-trivial transverse space
metric.
Then  there  may  exist  some special cases in which  such a model may remain
conformal to all orders once it is conformal
to  the leading order.  An example is  provided by
\eqn\twvec{ L = F(x) \[ \del u + 2S_i (x)  \del x^i\]\[\bd v
+ 2A_i (x) \bd x^i\] +
       \del x_i \bd x^i + \a' {\cal R} \phi(x) \ .  }
This model is related by $u$-duality  to the  $u$-independent case of the
`non-chiral'   generalization  of the $K$-model  \mokas\ with two
non-vanishing vector couplings (the relation of the functions  is $
F= K\inv(x) , \ S_i= \bar A_i (x) , \ A_i = A_i (x), \ \p=\p_0 + \ha \ln
F(x)$).
In the case when $S_i$ and $A_i$ have constant field strengths  \cobf,
the theory \twvec,   like \mokas,  can be shown to be conformally invariant
to all loop orders,
provided (cf. \coni)
\eqn\sonm{- \ha  \del^2 F\inv
+    {\cal F}^{ij} \bar {\cal F}_{ij} =0 \  , \ \ \ \ \p= \p_0 + \ha \ln F \ ,
\ \ \ \  \bar {\cal F}_{ij} = 2 \del_{[i} S_{j]}  \ .   }
The  proof  is a simple version of the  arguments used in the previous
subsections (in the special case of $S_i=-A_i$ it was given  already in
the Appendix B of \horts). Introducing  the sources and integrating  out $u$
and $v$ one obtains  the following effective
$x$-theory (cf. \mokaa, \mofak, \ggg)
$$ L_{eff} =
  \del x_i \bd x^i   -  F\inv (x) \del U \bd V
+  2 A_i (x) \del U\bd x^i  +   2S_i (x) \del x^i \bd V $$
\eqn\rtr{+ \
   \a'{\cal R}\p'(x) + X_i \del \bd x^i
    \ ,  }
so that if   $A_i$ and $S_i$ are linear in $x$  all  conformal anomaly
contributions come only from one-loop diagrams.

%%%%%%%%%%%%%%%%%%%%%%%%%%%%%%%%%%%%%%%%%
\newsec{Superstring and heterotic string solutions}
%%%%%%%%%%%%%%%%%%%%%%%%%%%

\def \op {\hat \omega_+}
\def \om {\hat \omega_-}

So far we have discussed exact classical solutions of the bosonic theory.
A generalization to the case of the closed superstring theory is
straightforward. The superstring  action is given by the
$(1,1)$ supersymmetric extension of the bosonic $\s$-model \mox\
(with $x^\m= (u,v,x^i)$    in \mox\ replaced by $(1,1)$ superfields $\hat X^\m
(z, \theta, \bar \theta)$ ).
Repeating the arguments of section 5
starting with the $(1,1)$ supersymmetric extension of \mox\
  $\  I_{(1,1)} = \int d^2 z d^2\theta  (G_{\m\n} + B_
{\m\n})(\hat X)  {\cal D} \hat X^\m  \bar {\cal D}
\hat X^\n \  $
and using  that the one-loop conformal invariance conditions are the same
as in the  bosonic case one finds  that
our exact   bosonic backgrounds  also represent
superstring solutions.
One can also  start with    the component representation
(here $\hat \omega^m_{\pm n\m} =\omega^m_{\ n\m} \pm \ha H^m_{\ n\m}$)
\eqn\onen{I_{(1,1)} =
\int d^2 z [ (G_{\m\n} + B_
{\m\n})(x) \del
 x^\m  \bd  x^\n  + \l_{Rm} (\delta^m_{n}\bd   + {\hat \omega}^m_{ -n\m}(x) \bd
x^\m )\l^n_R   }  $$
 + \l_{Lm} (\delta^m_{n} \del  + {\hat \omega}^m_{+n\m}(x)\del x^\m)\l^n_L
 - \ha {\hat R}_{+ mnpq} \l^m_L \l^n_L \l^p_R \l^q_R
]\ , $$
 write down  the fermionic part of the  action  explicitly with the help of
 (A.9),(A.16) and
directly integrate over the   left and right fermions.
One then finds that the only effect of the fermionic  contributions
on the effective bosonic $x^i$-theory is to cancel the
local $\del \ln F \bd \ln F$ term  coming from the bosonic
$u,v$-determinant.\foot{A simple test that this cancellation does take place
is provided by the observation that the two-loop $\b$-function
 must vanish  (in a ``supersymmetric"  scheme) in
the $(1,1)$ supersymmetric \sm \alvf, while the one-loop
induced  term $\del \ln F \bd \ln F$ term would contribute to the two-loop
conformal anomaly.}
Thus there is no need for a special adjustment  of a scheme
compared to  the  pure bosonic case (see also \horts).

As for the heterotic string solutions,
one approach is to start with
a closed superstring solution and  embed it into  a  heterotic
 string theory by
identifying the  generalized
Lorentz connection  $\hat \omega^m_{+n\m}$ (or $ \hat \omega^m_{-n\m}$) with  a
 Yang-Mills  background,
i.e.  by   rewriting the $(1,1)$ supersymmetric \sm  in the
$(1,0)$ (or $(0,1)$) supersymmetric heterotic \sm  form
\refs{\cand,\call,\senn,\huwi}.
For this to be possible,
the holonomy group of the generalized connection $\op$ (or $\om$)
 should be a subgroup  of  the heterotic string gauge group.
In general, such embedding is problematic  for solutions
with a curved space-time  (i.e. with a non-trivial
time-like direction) since the holonomy is then (a subgroup of) a non-compact
Lorentz group $SO(1,D-1)$ while the heterotic gauge group
should be compact on unitarity grounds.\foot{A special case of this
 was pointed out in
\kalor. Notice that if the  gauge group is non-compact,
at least one of the internal  fermions has a   negative norm but
 (compared to the $(1,1)$  supersymmetric superstring case) there is no extra
local world sheet superconformal symmetry to gauge it away \barss.}
In fact, as shown in Appendix A.2, the holonomy groups
of $\op$ and $\om$ for generic
chiral null models  are non-compact  (except for the case of the
plane wave background \yyys\ when  the holonomy of $\op$ is $SO(D-2)$)
and thus cannot be embedded into $SO(32)$ or $E_8\times E_8$.

%%%%%%%%%%%%%%%%%%%%%%%%%%%%%%%%%%%%%%%%%
\subsec{Exact heterotic string solutions}
%%%%%%%%%%%%%%%%%%%%%%%%%%%

One should thus try  a more direct  approach. As indicated above,
given  a bosonic string theory,  there exist, in principle,
 two  possible ways to construct a heterotic string theory
depending on whether the  ``right"  or ``left" parts of the bosonic coordinates
are supersymmetrized, i.e. on whether one  considers  a $(1,0)$ or $(0,1)$
supersymmetric world sheet theory. The two  heterotic theories are related by
interchanging left- and right- movers
in the vertex operators,  and, in general,  are
inequivalent. The fermionic parts of the heterotic \sms
corresponding to the two theories
depend   on $\om$ and $\op$ respectively.\foot{In particular, the \sm
$\b$-functions and low-energy effective actions
corresponding to the two theories  are related by simply changing the sign
of $B_{\m\n}$ (the effective actions of bosonic or supersymmetric string
theories  are invariant under  $B_{\m\n}\ra - B_{\m\n}$  since these theories
are invariant under  world sheet parity transformation).
That implies, e.g.,  that  the  ``right"  and  ``left"
heterotic extensions of a
 bosonic background which is chiral
(i.e. which  distinguishes between  left and right, e.g.,
having $B_{\m\n}\not=0$)
will be inequivalent.}
In what follows we shall concentrate on the standard
$(1,0)$ (or ``right") theory  since it  turns  out that the  $(0,1)$ (or
``left") theory
does not have chiral null models as exact solutions.

The action of the $(1,0)$ heterotic \sm is given by
 (we ignore the ``internal" fermionic part with a possible  gauge field
background)
\eqn\hett{ I_{(1,0)} =
\int d^2 z d\theta (G_{\m\n} + B_
{\m\n})(\hat X) {\cal D}
\hat X^\m  \bd \hat X^\n } $$
= \int d^2 z \[ (G_{\m\n} + B_
{\m\n})(x) \del
 x^\m  \bd  x^\n  + \l_{Rm} (\delta^m_{n}\bd   + {\hat \omega}^m_{ -n\m}(x) \bd
x^\m )\l^n_R \]\ . $$
The  (1,1) superstring \sm action \onen\ can be  formally  obtained from \hett\
  by
adding the internal left fermionic part coupled to   the  gauge field
background  equal to $\op$.

 Thus $\om$ appears in the  fermionic part of the \sm action \hett\ (and also
in the leading-order space-time supersymmetry transformation laws).
The $\b$-functions and the effective action  $S$ of this theory will
depend on $\om$ {\it but  also }  explicitly on the curvature $R$ of $G_{\m\n}$
and the  antisymmetric tensor field strength $H$.
 The \sm anomaly  will also
naturally involve $\om$. However,
 since the form of the anomaly  is ambiguous (scheme dependent)
\refs{\hulla,\huto}
  it can be arranged  so that  it will be
$\op$  that will enter the
anomaly relation  as well as  the ``anomaly-related" terms in the effective
action   (this, in fact,  is a common assumption,  see e.g.
\refs{\bergr,\chs}).
 It should be emphasized
that  there is {\it  no} unambiguous definition of such ``anomaly-related"
terms   since
$S$ is scheme dependent  and,  in general, cannot be  represented only in terms
of $\op$. There are always other $H$-dependent terms which are not expressed in
terms of the generalized curvature of $\op$ (so that one can equally well use
$\om$ in place of $\op$ at the expense of modifying the rest of the
terms).\foot{$O(\a')$-terms in the heterotic string effective action were
computed in \gros\ and \metts\ starting from the string $S$-matrix. As was
shown in \metts,
there exists a scheme in which the $\a'$-term (its part which is not related to
Chern-Simons modification of the leading-order $H^2$-term) in the heterotic
string action is the same as in the bosonic string one up to an extra  overall
factor of 1/2.
The same result was obtained from the analysis of the 3-loop conformal anomaly
of the heterotic \sm \keto.}

Let us now show that  our bosonic
solutions are exact solutions of the  heterotic string theory
without  any  extra  gauge-field background:
the direct $(1,0)$ supersymmetric extension
of the bosonic $\s$-model \mox\  is conformally invariant
if  the bosonic model is conformal.
The fermionic part of the action \hett\ does not actually
contribute to the conformal anomaly.
This follows from  the special ``null"
holonomy property of $\om$: according to Appendix A (see (A.16))
the only non-vanishing component of the generalized
Lorentz connection $\om$ is ${\hat \omega}_{-\hat u \hat i \m}$
($\hat u, \hat v, \hat i$ are tangent space indices).\foot{This property of
$\om$ is also responsible for the ``one-half" extended space-time supersymmetry
of our  bosonic backgrounds when they are embedded into $D=10$
supergravity as shown for the special cases  of  the (generalized) FS
in \refs{\gibb,\wald} and  for the
$F=1$ and $K=F\inv$  cases in \refs{\bergsh, \kallosh} (our notation for $\om$
and $\op$ are opposite to that of
\refs{\bergsh, \kallosh}).  The general chiral null model also has
unbroken spacetime supersymmetry, at least to leading order in $\a'$.
It should be possible to address higher
order corrections to the spacetime supersymmetry transformations
for this model
 in the worldsheet approach using  Green-Schwarz superstring action
in a light-cone type gauge (cf. \wald).
}
The non-trivial fermionic terms in \hett\ are thus
 given by
\eqn\ferm{ L_{(1,0)} (\l_R) = \l^{\hat u}_R \bd \l^{\hat v}_R
+ \l^{\hat i}_R \bd \l^{\hat i}_R
+  {\hat \omega}_{-\hat u \hat i \m}(x)
 \bd x^\m \l^{\hat u}_R \l^{\hat i }_R
 \ . }
The ``null" structure of the coupling
implies that integrating out fermions does not
produce a non-trivial contribution to
the $x^\m$-theory which remains conformally
invariant. There is an obvious similarity with
integrating out $u$ and $v$ in the  bosonic theory (cf. Section 5).

Thus we do not need a non-trivial gauge field background
to promote our bosonic solutions to heterotic ones.
We conclude that, for example,  the exact $D=5$ bosonic  solutions
 \yyy\
 are also heterotic string solutions
and so  are  their four dimensional `images' \didii.
In particular, the  $D=4$ extremal electric black holes discussed
 in Section 4.2 are thus  exact  heterotic string solutions \hrt\
without  any extra gauge field background.

Let us   compare    the above conclusion  with
the perturbative result  for the two-loop $\b$-function
of the heterotic $\s$-model.  Let us consider the
``non-anomalous" $\a'$ contribution  to the metric $\b$-function
$\b^G_{\m\n}$ (i.e. we shall ignore  other non-covariant $\a'$-corrections
which
modify the one-loop $H^2$-term by the Chern-Simons terms). The contribution
 of the  fermions $\l_R$ is essentially the same form as the standard two-loop
``$F^2$"-term that comes from the internal fermionic sector $\psi_L$ \call\
except for the fact that
 the gauge field is represented by the connection $\om$ \ross. Thus
\eqn\bett{ ({\b^{G (2)}_{\m\n}})_{(1,0)} =({\b^{G (2)}_{\m\n}})_{0}
-\fourth \a'{\hat R}_{- mn \l\m}{\hat R}^{mn \l}_{-  \ \ \ \n} \ , }
where $({\b^{G (2)}_{\m\n}})_{0}$ is the bosonic contribution. There exists a
special chiral ``right" scheme  in which the latter is given by  \metts
\eqn\bettb{ ({\b^{G (2)}_{\m\n}})_{0} =\fourth  \a'\( 2
{\hat R}_{- \ \ (\m}^{\a\b \l}{\hat R}_{-\n) \a\b \l}
 -  {\hat R}_{- \ \ \ (\m}^{  \b\l\a}{\hat R}_{-\n) \a\b \l}
+
   {\hat R}_{- \a (\m\n) \b}  H^{\a\r\s}H^\b_{\ \r\s} \)\ . }
As follows from (A.9)  $ {\hat R}_{- mn \l\m}{\hat R}^{mn \l}_{- \ \ \ \n}$
(i.e.
the fermionic contribution) indeed vanishes for our backgrounds.
As for \bettb, it also vanishes when $F=1$ but in general one needs to choose
a different  scheme to avoid $\a'$-corrections (see \horts).

Given the scheme dependence of the $\b$-function, in the {\it heterotic } \sm
context  there may exist
a scheme in which
 the bosonic contribution to the  \sm $\b$-function  \bett\
can be put in the following ``left-right symmetric" form
 \eqn\betb{ ({\b^{G (2)}_{\m\n}})_{0} = \fourth \a' \(
 {\hat R}_{+ mn \l\m}{\hat R}^{mn \l}_{+ \ \ \ \n}
+  {\hat R}_{- mn \l\m}{\hat R}^{mn \l}_{- \ \ \ \n} \) \ . }
Including the gauge field contribution of the internal  left
fermions  the
heterotic \sm $\b$-function corresponding to this  ``symmetric"  scheme
then is given by
\eqn\betbf{ ({\b^{G (2)}_{\m\n}})_{(1,0)} = \fourth \a'
 {\hat R}_{+ mn \l\m}{\hat R}^{mn \l}_{+ \ \ \ \n}  - \fourth \a' F_{IJ\l\m}
(V) F^{IJ\l}_{\ \ \   \ \n }(V)\ . }
This expression  is consistent with the  expectation that
the two-loop $\b$-function should vanish once we identify the gauge field
$V_\m^{IJ}$ with $\op$ since then the heterotic \sm becomes identical to the
(1,1) supersymmetric model \onen.
The two-loop contribution \betbf\ with $V_\m=0$ does not  vanish
for our backgrounds
even in the  simplest plane-wave case $F=1$.
 As already mentioned above,
in general, we
cannot  make it vanish by the identification $V=\op$  since the holonomy group
of $\op$ is non-compact. Thus in this scheme our solutions will be modified by
higher-order $\a'$ corrections.

In the  special case of $F=1, K=0, A_i=A_i(x)$,
 the only non-vanishing component of $\op$ is
${\hat \omega}_{+  \hi\hj u}= -  {\cal F}_{ij}  $ and
one finds that $ {\hat R}_{+ mn \l\m}{\hat R}^{mn \l}_{+ \ \ \   \n}$ in
\betbf\ has non-vanishing $uu$-component equal to $ (\del_k {\cal F}_{ij})^2$.
If we set $V_\m=0$ and start with the leading-order solution
$K=0, \ \del_i {\cal F}^{ij}=0$ then $K$ receives the $\a'$-correction $K_1$
satisfying
(cf. (A.27))
$-\ha \del^i\del_i K_1 + \fourth \a' (\del_k {\cal F}_{ij})^2  =0$,
i.e. $K_1= \fourth  \a' ({\cal F}_{ij})^2 $.  Such modification can be
thought of as a local field redefinition corresponding to the transformation
from
the chiral ``right" scheme \bett,\bettb\  where $K_1=0$ to the ``left-right
symmetric" scheme \betbf.\foot{ This redefinition of $G_{uu}$ could be
 thought of as induced by
$G_{\m\n}'=G_{\m\n} +  \fourth \a' H_{\m\l\r} H_\n^{\ \l\r}$.
It  may also be related to   the { non-covariant} redefinition
$G_{\m\n}'=G_{\m\n} +  \fourth \a' ({\hat \omega}^{mn} _{+  \  \m}{\hat
\omega}_{+  mn \n} -  V^{IJ}_\m V_{IJ\n})$
used  in  \huto\ in order to preserve world sheet supersymmetry (there is only
$\op\op$-term   if $V=0$ and the whole redefinition is  trivial
 if $V=\op$). }

Since in this exceptional  case  the holonomy
of $\op$ is compact ($SO(D-2)$), there is also an alternative option
to  introduce the gauge field background $V^{IJ}_\m$ equal to  $\op$
and in this way cancel the higher order correction.
This  was suggested  in \bergsh\  where \betbf\ was assumed
to be the  form of the $\a'$-correction in the heterotic string equation
of motion.\foot{While the effective action considerations in \refs{\bergsh,
\kallosh} are not sufficient
to demonstrate the exactness of the  solutions to all orders in $\a'$ since
they were ignoring ``anomaly-unrelated" terms (in particular,  no explanation
was given why these backgrounds are superstring solutions, i.e. why the
corresponding  $(1,1)$ supersymmetric \sm should be conformally invariant),
this is  possible within  our  direct
world sheet  approach.  Although the approach of \refs{\bergsh,
\kallosh,\kalor}
is incomplete, our present work was much  motivated and influenced by these
interesting  papers.}
As we have mentioned, the idea of embedding of $\op$ into the gauge group does
not have
consistent generalizations to other  cases except the one with $F=1, K=0,
A_i=A_i(x)$
%\foot{A related  observation  was  made  in \kalor.}
so that we disagree with the suggestion  of \refs{\bergsh, \kallosh}
that   $F=1$ and $F=K\inv$  models  are exact  heterotic string solutions only
when  supplemented by   a gauge field background.
The need to introduce  a non-trivial gauge field background in
\refs{\bergkal,\kallosh} was  caused  by having implicitly taken  the $\a'$
term in the effective action  in a specific
``symmetric scheme"    (in which   $\op$  appears in the  ``anomaly-related"
terms).
As we have explained above,  the form of $\a'$-corrections is scheme dependent
and in the natural chiral scheme there is no need for an extra gauge field
background at all.

 The plane wave model  \yyys\ with   $F=1, K=0, A_i=A_i (x)$  and the  gauge
field background $V^{ij}_u = {\hat \omega}_{+  \hi\hj u }= -  {\cal F}_{ij} $
is equivalent to the $(1,1)$ supersymmetric superstring \sm
and  thus   represents an exact  solution according to the discussion  at the
beginning of this section.
It is instructive to see explicitly why the resulting model remains
conformally invariant: the fermionic terms   now are ($\psi_L^m$ are internal
fermions; see also (A.9),(A.16))
\eqn\fermr{ L_{(1,1)}(\l_R,\psi_L) = \l^{\hat u}_R \bd \l^{\hat v}_R
+ \l^{\hat i}_R \bd \l^{\hat i}_R
+  {\cal F}_{ij}(x)
 \bd x^j \l^{\hat u}_R \l^{\hat i }_R } $$
 +  \psi^{\hat u}_L \del \psi^{\hat v}_L
+ \psi^{\hat i}_L \bd \psi^{\hat i}_L
-   {\cal F}_{ij}(x)
 \del u \psi^{\hat i}_L \psi^{\hat j }_L
  -  \ha \del_i {\cal F}_{jk} (x) \l^{\hat u}_R \l^{\hat i }_R\psi^{\hat j}_L
\psi^{\hat k }_L  \ . $$
Integration over $\l^{\hat v}_R $ `freezes' out $\l^{\hat u}_R$,
while the  term ${\cal F}_{ij}(x)
 \del u  \psi^{\hat i}_L \psi^{\hat j }_L$ does not produce new divergencies
in the $uu$-direction since the total action does not contain local
 $\bd u$-couplings (cf. \xxx).

Finally, let us  consider
 the (0,1) (``left") heterotic theory. Here the superstring fermions are
coupled to $\op$. Since according to (A.17) $\op$ has  general holonomy,    one
should expect non-trivial fermionic contributions to the conformal anomaly.
The gauge field background cannot be consistently introduced
since the (abelian) holonomy group of $\om$ is ``null" (non-compact).
 The corresponding leading-order
solutions thus should  have corrections to all orders in $\a'$.
Given that $\op$ (which in this theory appears also in the space-time
supersymmetry  transformation laws) is of generic form, one should not also
expect to find Killing spinors, i.e. a residual supersymmetry.

%%%%%%%%%%%%%%%%%%%%%%%%
\subsec{Extended world sheet supersymmetry}
%%%%%%%%%%%%%%%%%%%%%%%%%%%%%
%%%%%%%%%%%%%%%%%%%%%%%%%%%%%%%%%%%%%%%%%%%%%%%%%%%%%%%%%%%%%%%%%%%%%%%%%%%%%%%
 It is clear that the   abelian gauge fields of the four dimensional
 solutions  \didii\ have  a
Kaluza-Klein and not a heterotic Yang-Mills origin.
In general, given a $D=4$ leading-order bosonic  background, its  embedding
into the heterotic string theory is not unique.
The  embeddings of extremal $D=4$ dilatonic black holes
in which the $U(1)$ gauge fields have
 a Kaluza-Klein (i.e. $N=4$ supergravity)
and not  a  heterotic
Yang-Mills origin
have extended (e.g. $N=2, \ D=4$) space-time supersymmetry \susy.
Since  our general bosonic $D\leq 10$  backgrounds
have extended space-time  supersymmetry when embedded into $D=10$ supergravity
theory \refs{\gibb,\wald,\bergsh,\kallosh,\kall}
one  may
also try to envoke  supersymmetry to argue  that they are exact superstring
solutions.

In fact, the presence of extended space-time supersymmetry suggests
(cf. \refs{\sss,\chs}) that the
corresponding $(1,0)$ supersymmetric \sms may have extended world sheet
supersymmetry.
If the latter supersymmetry is large enough,
one may use the fact that there exists a scheme in which the $(4,n)$
supersymmetric \sms are conformal to all orders
\sus.

 In contrast to our approach  described in Section 5 and in the previous
subsection, any
 argument based on extended world sheet supersymmetry is bound to have a
limited applicability.
The standard
discussions of extended world sheet supersymmetry apply  to the case of
Euclidean target space signature. To have $(2,n)$ supersymmetry the dimension
$D$ must be even; the  $(4,n)$ supersymmetry is possible only when $D$ is
multiple of 4.
 Most of our models (e.g. all with odd  spacetime dimension) do not admit
extended world sheet supersymmetry  since they do not admit a complex structure
when analytically continued to Euclidean signature.

The generic  chiral null model \mox\
does not have a natural analytic continuation  with  a real Euclidean target
space
metric. For example, if one analytically continues $u+v$ keeping $u-v$ real,
so that $u$ and $v$ become
complex conjugates ($v=\bar u$),   then   the metric is no longer real
unless $K$ and $A_i$ in \mox\  are taken to be zero.
There may exist a well-defined Euclidean analog of \mox\  for some
special choice of $A_i$ but  we shall  ignore this possibility  for simplicity.
In the special case of the \FM  \mof\  one gets a  real  action
 on the Euclidean world sheet
 (but thus a complex string action in the Minkowski world sheet signature case)
\eqn\rea{ L= F(x) \del u \bd \bar u + \del x_i \bd x^i  +  \a'{\cal R}\p (x) \
. }
The corresponding
 Euclidean metric
$\  ds^2 = F(x) dud\bar u+ dx_idx^i  \ $
 is real but
the antisymmetric tensor   is  imaginary.
If the dimension is even,
$D=2N$,
 the metric  is   hermitian.
  Replacing $x^i$ by a set of complex coordinates $w_s$ ($s=1,...,N-1$)
the metric and the antisymmetric tensor are
 \eqn\des{ ds^2 = F(w,\bar w) dud\bar u  + dw_sd\bar w_{ s}\ , \ \ \  \
B_{u\bar u} = \ha F\ , \ \ \  \
 H_{s u\bar u}= \ha \del_s  F \ .  }
The corresponding  $(1,0)$ supersymmetric
\sm admits  $(2,0)$ extended supersymmetry.
This is clear  from  the comparison with the conditions on geometry implied by
$(2,0)$
supersymmetry \huwi,  as reviewed, e.g., in  \bonn\ (for some earlier
discussions of related complex geometries, see  \refs{\suhol,\sev}). Provided
$\del^2
F\inv =0$  the generalized connection  with torsion  has special (not $U(N)$
but $SU(N)$) holonomy and satisfies  the generalized quasi Ricci flatness
condition (see Appendix A)
 \eqn\ricc{ {\hat R}_{\m\n} = {\hat D}_\m V_\n\  , \ \ \  \   \ V_\m = -\del_\m
\ln F\  . }
If the dimension $D$ is a multiple of four, i.e. $N=2N'$,
 a $(2,0)$ supersymmetric \sm may admit $(4,0)$ extended supersymmetry. In
fact, the Euclidean \FM \rea\ does have it, as   is clear again from the
comparison with the expressions in \bonn.
In particular, the  holonomy of the generalized connection  is an  $Sp(N') $
subgroup of $ SU(N)$.\foot{
In the simplest case of $D=4$  and $F=F(|w|) $ the  metric becomes conformal
to a K\"ahler metric, cf. \refs{\chs,\bonn}.
}
Given that $(4,0)$ supersymmetric \sms
are conformally invariant to all orders (in a properly chosen scheme) \sus\
we get (for $D=4N'$)
an independent  proof of the fact   that the \FM corresponds to an exact
solution
of heterotic string theory.\foot{The conclusion about  extended supersymmetry
of  the \FMS is consistent with the fact that some  of them  correspond to
special   nilpotently gauged WZW
models  \klts.  The latter  are formulated essentially in terms
of the WZW model on a (maximally non-compact) group $G$ and thus
their Euclidean versions  should  admit  $(2,0)$ or $(2,2)$
supersymmetry  \refs{\sev,\hus}.}

It should be stressed again that our explicit
 proof given in \horts\ and in the
present paper is more direct and  applies  for  any $D$ as well as
to a more general class of models \mox.
In general,  a relevant property which is important for the proof  of exactness
 is the special   holonomy
of the generalized connection  with torsion and
 not an extended supersymmetry (which is  just a consequence of the special
holonomy  under certain additional conditions like  existence of a  complex
structure).\foot{ A somewhat related  remark  was  made in \howe, where it was
 pointed  out   that   since the   \sm on a  Calabi-Yau space has a special
holonomy it  thus has an extra infinite-dimensional non-linear classical
symmetry. That symmetry (if it were not anomalous  at the quantum level)
would  rule out all higher-loop corrections to the  $\b$-function  \howe.
In our case, the analogous symmetry is linear and  is the affine symmetry
generated by  the null chiral
 current. }

%This is a  general point against giving to much sense to the fact that
%in some cases there is a connection between space-time susy and
%extended world sheet susy etc.

%%%%%%%%%%%%%%%%%%%%%%%%%%%%%%%%%%%%%

\subsec{Relation to  other $D=4$ heterotic solutions}
%%%%%%%%%%%%%%%%%%%%%%%%%%%%%%%%%%%%%%%%%%%%%%%5
What about non-supersymmetric solutions of $D=4$ heterotic string theory?
For example,  the charged dilatonic black hole  may be considered as   a
non-supersymmetric leading-order solution \gar\ of the  $D=4$ heterotic string
theory
with the charge
being that of the $U(1)$ subgroup of the  Yang-Mills  gauge group.
This solution must have an  extension to
higher orders in $\a'$  which, in general,  may not be the same
as the  above  supersymmetric `Kaluza-Klein' solution obtained
by dimensional reduction from $D=5$.
Even though the leading-order terms in the
 compactified (from $D=5$ to $D=4$) bosonic
 string theory and $D=4$ heterotic string theory
with a $U(1)$ gauge field background
look the same, the $\a'$-corrections are  different, so
that  our bosonic  result  does not  automatically imply  that the extremal
electric black hole  considered as a  $D=4$  heterotic string solution is also
exact.
In fact, it is known that the non-supersymmetric extremal magnetic solution
of the $D=4$ heterotic string has  $\a'$-corrections \nats.
The same is likely to be true for the
non-supersymmetric extremal electric solution.

To explain this difference  between  ``supersymmetric" and ``nonsupersymmetric"
solutions  it is useful to consider  the   space-time effective action
approach.
Our  exact  $D=4$ solutions  \dididi\ obtained by  dimensional reduction
are  actually  $D=5$   bosonic or heterotic string  solutions.
This means that there exists a  choice of (five dimensional) field
redefinitions in which the
$D=5$ effective
equations evaluated on our background
contain no $\a'$ corrections. As shown in  section 3, the dimensional
reduction of the $D=5$ action includes $two$ gauge fields (as well as an
extra scalar modulus field). Even though these two gauge fields
are equal for our solution \dididi,  the general field redefinition
treats them independently. In contrast, the $D=4$ non-supersymmetric
heterotic action contains a single gauge field and thus a smaller
group of field redefinitions. Thus the fact that nontrivial $\a'$
corrections  inevitably arise in this case (for the magnetically charged
black hole \nats\ and, most likely,   for the electrically  charged case as
well)
does not contradict our claim that the supersymmetric electrically
charged solution obtained from dimensional reduction is exact.

In general, the starting point is the $D=10$ heterotic string
with the leading-order term in the effective action being
represented by the $N=1, \ D=10$ supergravity coupled to $D=10$
super Yang-Mills theory. Compactified on a  6-torus,
this effective action   becomes that of $N=4, D=4$ supergravity
coupled to a number of abelian $N=4$ vector multiplets and
$N=4$ super Yang-Mills.  The simplest  charged dilatonic black hole solution
may be embedded in this theory in several inequivalent  ways, depending
on which vector field(s) is kept non-vanishing.
The dependence of higher-order $\a'$-terms in the full effective action
on different vector fields is different, so it should not be surprising that
the solutions that  happen to coincide at the leading-order level may turn out
to receive different $\a'$-corrections.

%%%%%%%%%%%%%%%%%%%%%%%%%%%%%%%%%%%%%%%%%%%%%%%%%%%%%%%%%%%%%%%

Finally,
let us note that  it may be possible
 to utilize  some of our  $D>4$ exact bosonic solutions to construct other
$D=4$ heterotic string solutions.\foot{To establish a relation between
heterotic and bosonic models one can use the following strategy: start with a
leading-order heterotic string solution, write down the corresponding heterotic
\sm and then try to bosonize it to put it in a form of a  bosonic \sm for which
it may be possible  to prove  the conformal invariance directly.}
The idea is to start with an exact higher dimensional bosonic solution and
then fermionize the `internal' coordinates in an appropriate way to
obtain a heterotic $\s$-model. A similar method was used   in
\gps\ to find
the heterotic solution representing
 a $D=2$ monopole theory
(which was related to the throat limit of the $D=4$ extreme
magnetically charged black hole)
 and in \jon\ to describe a non-trivial
throat limit of the $D=4$ dilatonic Taub-NUT solution \refs{\kall,\jons}.

%%%%%%%%%%%%%%%%%%%%%%%
\newsec{Concluding remarks}
%%%%%%%%%%%%%%%%%%%%%%%%%

To obtain exact solutions in  string theory,  it is rather hopeless to  start
with the
field equations expressed as a power series in $\a'$, and  try  to solve
them explicitly.  One must first make an ansatz which simplifies
the form of these equations. We have studied such an ansatz, the
chiral null models \mox, and shown that they have the property
that there exists a scheme in which the leading order
string solutions  are exact. This generalizes a number of previous results.
The chiral null models include the plane wave type solutions and the
fundamental string background which were previously shown to be exact. But
as we have seen,
they also include,  e.g., the solution describing traveling waves along the
fundamental string,
and,  after  a dimensional reduction, the extremal electrically charged
dilaton black holes and the dilaton IWP solutions.
Moreover, there are interesting solutions describing magnetic field
configurations.
It is rather surprising
that such a large class of leading-order solutions
turn out to be exact in bosonic, superstring $and$ heterotic string theories.

One can, in fact, turn the argument around. Even the leading order string
equations (analogous to Einstein's equations) can be rather complicated when
the dilaton and antisymmetric
tensor are nontrivial. By choosing an ansatz at the level of the string
world sheet action  which yields simple equations for the \sm  $\b$-functions,
one can easily find new solutions of even the leading order equations. The
chiral
null models are an example of this. Some of the solutions we have discussed,
e.g. \didii\ with a general $K$, appear to be new.

However, it is clear that
not all of the solutions of the leading order equations can be obtained
from chiral null models. The chiral coupling, which is an important feature
of these models, leads to a no-force condition on the solutions, and the
possibility of linear superposition. This  happens  only for a  certain
charge to mass ratio which  typically characterizes  extreme black holes
or black strings. Furthermore, we have obtained only  four dimensional
black-hole type
solutions with {\it electric}  charge. Extreme magnetically charged black holes
do not appear  to be
described by chiral null models.

We have  considered examples of
chiral null models with a flat transverse space.
As we have remarked, it is straightforward to extend this class of models to
any
transverse space which is itself an exact conformal field theory.
It may be interesting to explore the new solutions
(with non-trivial mixing of ``space-time" and ``internal" directions)
 which can be obtained in
this way.

An important open problem is to study string propagation in the backgrounds
discussed here. This will improve our understanding of the physical
properties of these solutions in string theory.

%%%%%%%%%%%%%%%%%%%%%%%%%%%%%%%%%%%%%%%%%%%%%%%%%%%%%%%%%%%%%%%%%%%
\newsec{Acknowledgements}
%%%%%%%%%%%%%%%%%%%%%%%%%%%%%%%%%%%%%%%%%%%%%%%%%%%%%%%%%%%%%%%%%%%%
We would like  to thank  G. Gibbons,   R. Kallosh and A. Strominger for  useful
discussions.   G.H. was supported in part by NSF Grant PHY-9008502.
 A.A.T. is grateful to CERN Theory Division for hospitality while this paper
was  completed and  acknowledges  also the support of PPARC.

%%%%%%%%%%%%%%%%%%%%%%%%%%%%%%%%%%%%%%%%%%%%

%%%%%%%%%%%%%%%%%%%%%%%%%%%%%%%%%%%%%%%%%%%%
\def \H {{\cal H}}

\appendix{A}{Geometrical quantities for the chiral null  model}
%%%%%%%%%%%%%%%%%%%%%%%%%%%%%%%%%%%%%%%%%%%%%%%%%%%%%%%%%%%%%%%%
%%%%%%%%%%%%%%%%%%%%%
\subsec{Generalized connection}
%%%%%%%%%%%%%%%%%
The classical string equations for a
\sm
\eqn\mok{L=  C_{\m\n} (x) \del x^\m \bd x^\n \ , \ \ \ \ \
 C_{\m\n} \equiv  G_{\m\n} +  B_{\m\n} \ , }
 are naturally expressed in terms of the generalized connection with torsion
\eqn\conn{ \del\bd x^\l + \hG^\l_{-\m\n} (x) \del x^\m \bd x^\n =0\  \ \
{\rm or} \ \
\del\bd x^\l + \hG^\l_{+\m\n} (x) \del x^\n \bd x^\m=
0 \ , \ \ }
\eqn\con{ \hG^\l_{\pm\m\n}  =\G^\l_{\m\n} \pm
 \ha {H^\l}_{\m\n} \ , \ \ \ \hG^\l_{-\m\n}= \hG^\l_{+\n\m}=
 \ha G^{\l\r}( \del_\m C_{\r\n} + \del_\n C_{\m\r} - \del_\r C_{\m\n}) \  . }
In the case of our model \moxa\ $\  x^\m = ( u,v, x^i)$ and
\eqn\mod{ G_{uv}= \ha F \ , \ \  G_{ui}=  F A_i \ , \ \  G_{uu } = F K \ , \ \
G_{vi} = 0 \ , \ \   G_{vv}=0 \ , \ \ G_{ij} =\d_{ij} \ ,  }
$$ G^{uv}= 2 F\inv\    , \ \ \ \  G^{ui}=  G^{uu } = 0  , \ \ \
G^{vi} = - 2 A^i   \ , $$ $$  \ \  G^{vv}=4 ( A_iA^i - F\inv K)\   ,
 \ \ \ \ G^{ij} =\d^{ij} \  , $$
\eqn\modw{ C_{uv}=  F \ , \ \  C_{vu}=0\ ,  \ \  C_{ui}= 2 F A_i \ , \ \
C_{iu} =0 \ , \ \  C_{uu } =  F K \ , \ \ } $$
C_{vi}=C_{iv} = 0 \ , \ \  C_{vv}=0 \ , \ \ C_{ij} =\d_{ij} \ . $$
We shall use the following definitions
\eqn\dee{ h (x) =\ \ha \ln F(x) \ ,    \ \ \
{\cal F}_{ij} = \del_i A_j - \del_j A_i \  , \ \ \       K  =  K (x,u)  \
, \ \  A_i  =  A_i(x,u)  \ .  }
 The corresponding components of the connection are ($\hG^\l_{+\m\n}=
\hG^\l_{-\n\m}$)
\eqn\chr{ \hat\Gamma^u_{-ui}= 2\del_i h \ , \ \  \ \ \ \hat\Gamma^u_{-ij }
=\hat\Gamma^u_{- vj}=
\hat\Gamma^u_{-\m u} =\hat\Gamma^u_{-\m v}= 0\ ,    }
 $$
\hat\Gamma^i_{-uv}= - F\del^i h \  , \ \  \ \
\hat\Gamma^i_{-uu} = F\del_u  A^i - \ha \del^i (FK ) \  ,  $$ $$
\hat\Gamma^i_{-uj} = - A_j \del^i F    -  F {{{\cal F}^i}_j} \   ,  \ \  \ \
\hat\Gamma^i_{-v\m }= \hat\Gamma^i_{-j\m  } =0  \  ,  $$
 $$
\hat\Gamma^v_{-iv}= 2\del_i h \ , \ \
\hat\Gamma^v_{-ij } = 2 \del_i A_j  + 4 A_j \del_i h \ , \ \ \ \
\hat\Gamma^v_{-iu} = \del_i K + 2K\del_i h  \ ,  $$
$$
\hat\Gamma^v_{-ui} = \del_i K - 2 K \del_i h
+ 2A_i A^j \del_j F - 2F A^j  {\cal F}_{ij}  \ , \ \ \ \
\hat\Gamma^v_{-uv} = A^i\del_i F \ , \ \  $$ $$
\hat\Gamma^v_{-uu} =  \del_u K - 2 F A^i \del_u A_i + A^i  \del_i (FK)
\ , \ \ \ \
\hat\Gamma^v_{-v \m }=0 \ . $$
It is straightforward to compute the  curvature tensor
 corresponding to $\hG^\l_{\pm \m\n}$ (note that the torsion here is a closed
 form)
\eqn\tors{ {\hat R}^\l_{\pm \m\n\r}= {\hat R}^\l_{\  \m\n\r} (\hG^\l_{\pm
\m\n})\ ,
\ \ \  \ \   {\hat R}_{- \l\m\n\r}=  {\hat R}_{+ \n\r\l\m} \ . }
We get
\eqn\cur{ {\hat R}^u_{- \m\n\r} =0  \ , \ \ \
{\hat R}^i_{- j \n\r} =0\ , \ \  \  {\hat R}^\l_{- \m uv } =0 \ , \ \ \  {\hat
R}^\m_{- v  \n \r }=0
\ , } $$   {\hat R}^v_{- i v j } = 2F\inv {\hat R}^i_{- u j v} =
  - 2 \del_i \del_j h \ ,
\ \ \ {\hat R}^v_{- u v j } = -2 FA^i  \del_i \del_j h\  , $$
$$ {\hat R}^v_{- iuj}=2F\inv {\hat R}^i_{- uju} =
2\del_i \del_u A_j + 4 \del_i h \del_u A_j
-2K \del_i \del_j h - \del_i \del_j K - 2 \del_i h \del_j K \ , $$
$$   {\hat R}^v_{- ijk }= -2F\inv  {\hat R}^i_{- u jk } =
2\del_i {\cal F}_{jk} + 4 \del_i h {\cal F}_{jk}
+ 4 A_k \del_j \del_i h - 4 A_j \del_k \del_i h \ . $$
Note that   product of the curvatures
${\hat R}^{mn}_{-\ \l\r} {\hat R}_{- mn \m\n}$ vanishes.

%%%%%%%%%%%%%%%%%%%%%%%%%%%%%%%%%%%
\subsec{Special holonomy}
%%%%%%%%%%%%%%%%%%%%%%%%%%%%%%%%%%%%
The expressions for the curvature \cur\  reflect   holonomy properties of the
generalized connections
$\hG^\l_{\pm\m\n}$.
It turns out that the holonomy group of $\hG^\l_{- \m\n}$
is an abelian $(D-2)$ - dimensional ``null" subgroup
of  the Lorentz group $SO(1,D-1)$. The
 holonomy group of $\hG^\l_{+ \m\n}$
is not special for generic functions $F,K,A_i$
It  becomes  the Euclidean group
in $D-2$ dimensions when $F=1$ and  reduces further to
 its rotational
 subgroup $SO(D-2)$ when $F=K =1, A_i=A_i(x)$.

It is easy  to argue  that  a   special holonomy of the generalized connection
 $\hG^\l_{- \m\n}$
in \conn\ is a direct consequence  of  the presence of a chiral current
in the \sm \mok\ (for a related more general discussion see \howe\ and refs.
there).

 If one introduces the vierbeins and defines the following differentials
(or `currents')
\eqn\difi{ \tt^m = e^m_\m\del x^\m \ , \ \ \ \bt^m = e^m_\m\bd x^\m \ , \ \ \
 G_{\m\n} = \eta_{mn}  e^m_\m  e^n_\n  \ , }
where $\eta_{mn}$ is the tangent space metric,
then  the string equation \conn\ can be written in the form
\eqn\conne{ \bd \tt^m + {\hat \omega}^{m}_{- n \m}  \bd x^\m \tt^n  =0 \
\ \ {\rm or} \ \ \del \bt^m + {\hat \omega}^{m}_{+ n \m} \del x^\m \bt^n   =0 \
,}
where
${\hat \omega}^{m}_{\pm n \n}$ are  the generalized Lorentz  connections
\eqn\lor{ {\hat \omega}^{m}_{\pm n \n} = e^m_\l \hG^{\l}_{\pm \m\n}     e^\m_n
   +  e^m_\l  \del_\n   e^\l_n  \ . }
In the case of \moxa\ one  may choose (the tangent space indices take the
following values:  $m= (\hat u, \hat v , \hat i$))
\eqn\mmm{ \tt^{\hat u}  = F\del u \ , \ \ \  \
  \tt^{\hat v} = \del v +  K \del u + 2 A_i \del x^i \ , \ \ \
\tt^{\hat i} = \del x^i\  ,  }
so the Lagrangian \moxa\ takes the form
\eqn\zzzzz{L= \tt^{\hat u}\bt^{\hat v}
 + \tt^{\hat i} \bt^{\hat i} + \a' {\cal R} \p(x)\ .}
 Then the existence of the null $v$-isometry implying
$\bd \tt^{\hat u} =0$ tells us that
$ {\hat \omega}^{\hat u}_{-  n \m} =0 , \   $
i.e. that the  connection  ${\hat \omega}_{-}$  has a  reduced
holonomy.\foot{In the case
of the \FM  \horts\ one has  two null chiral currents ($u$ and $v$ are on an
equal footing) and so both `left' and
`right' connections  should   have  the same
properties.  Note, however, that our choice of vierbein in \mmm\
is not symmetric in $u$ and $v$ so an extra Lorentz transformation will be
needed to  relate $ {\hat \omega}_-$ to $ {\hat \omega}_+$.}

Defining the connection 1-forms ($\eta_{\hu\hv}=\ha, \ \eta_{\hi\hj}=
\delta_{\hi\hj}$)
\eqn\cof{  {\hat \omega}_{\pm  m n }=  \eta_{mp} {\hat \omega}^{p}_{\pm  n \m}
dx^\m
= -  {\hat \omega}_{\pm  nm} \ , }
we find $\  {\hat \omega}_{-  m\hv }=
{\hat \omega}_{-  \hi\hj }=0 , \ $ and
\eqn\oom{ \ \ \ \
{\hat \omega}_{-  \hu \hi }= \del_i h dv + (\ha \del_i K  - \del_u A_i+ K
\del_i h) du +
({\cal F}_{ij} + 2 A_j \del_i h) dx^j \  , }
\eqn\oomp{ {\hat \omega}_{+  \hu\hv }= \del_i h dx^i \ , \ \ \
{\hat \omega}_{+  \hi\hj }= - F {\cal F}_{ij} du \ , \ \ \  } $$
{\hat \omega}_{+ \hu \hi }=  (\ha \del_i K - \del_u A_i) du \ ,
\ \ \ {\hat \omega}_{+ \hv \hi }=  F \del_i h du \ .   $$
Since the algebra of the Lorentz group $SO(1, D-1)$
is generated by $M\equiv M_{\hu\hv},\  L_\hi\equiv M_{\hu\hi},\  R_\hi\equiv
M_{\hv\hi}, \ M_{\hi\hj}$
satisfying, in particular,
 \eqn\lorr{[M, M_{\hi\hj}]=0\ , \ \  [M, L_{\hi}] =
L_{\hi}\ , \ \   [M, R_{\hi}] =
- R_{\hi}\ , \ \
[L_{\hi}, R_{\hj}] = 2\delta_{\hi\hj} M +
 M_{\hi\hj}\  ,
} $$
 [M_{\hi\hj}, L_{\hk}] = 4 L_{[\hj}\eta_{\hi]\hk}\ , \ \ \
 [M_{\hi\hj}, R_{\hk}] =4 R_{[\hj}\eta_{\hi]\hk}\ ,
\ \   [L_{\hi}, L_{\hj}]=[R_{\hi}, R_{\hj}]=0\ ,
$$
we conclude that the holonomy group of ${\hat \omega}_{-}$
is equivalent to the non-compact abelian subgroup of the Lorentz group
 generated by $ M_{\hu \hi}$ (it is ``null",  having zero norm associated with
it).
The holonomy of ${\hat \omega}_{+ }$
is not special in general.\foot{In the absence of
torsion the
irreducible  holonomy groups (or ``special geometries") on non-symmetric spaces
have been classified \berg.  No systematic classification  seems to be known in
the torsionful case. We thank  J. De Boer and G. Papadopoulos for helpful
comments on this subject.}

Let us now consider some particular  cases.
When $F=\const$  we find that $ {\hat \omega}_{+ \hu \hv }={\hat \omega}_{+ \hv
\hi }=0$ and thus the holonomy algebra of ${\hat \omega}_{+   }$
reduces the Euclidean algebra generated by $M_{\hi\hj}$ and $M_{\hu \hi}$.
It reduces further to the algebra of  $SO(D-2)$ when $K =1, A_i=A_i(x)$ (i.e.
in the case of the model \yyys).

In the case of the  generalized FS solution
related to the  black hole type solutions \dididi\
we have $K=F\inv, \  A_i=A_i(x) $ in \yyy\
so that the non-vanishing components of the connections
are
\eqn\kkl{ {\hat \omega}_{-  \hu \hi }= \del_i h dv\  +
({\cal F}_{ij} + 2 A_j \del_i h) dx^j \  , }
$$
 {\hat \omega}_{+  \hu\hv }= \del_i h dx^i \ , \ \ \ {\hat \omega}_{+  \hi\hj
}= - F {\cal F}_{ij} du\ ,  $$
 $$
{\hat \omega}_{+ \hu \hi }=  -F\inv \del_i h du \ ,
\ \ \ {\hat \omega}_{+ \hv \hi }=  F \del_i h du \ .   $$
When $A_i=0$ the  holonomy algebra of $
 {\hat \omega}_{+}$
becomes
 the $2D-3$ dimensional non-semisimple subalgebra of the Lorentz algebra
generated by   $M, L_{\hi}$ and $ R_{\hi}$.

 A special holonomy is  known \refs{\suhol,\sev,\sus}   to be related to the
presence of extended  world sheet supersymmetry in the  supersymmetric
extensions of the $\s$-models.  In fact,  some of the models   \moxa\
(which, in  particular, admit a complex  structure)
have extended supersymmetry (see Section 6).
Let us note also that special holonomy does not guarantee,  by itself,
conformal
invariance  since for that the dilaton is crucial as well.
Still,    it is  related (in  a proper renormalization scheme) to the  on-shell
finiteness of  our models on a flat world sheet.

%%%%%%%%%%%%%%%%%%%%%%%%%%%%%%%%%%%%%%%%%%
\subsec{ Parallelizable spaces and connection to WZW models based on
non-semisimple groups}
%%%%%777%%%%%%%%%%%%%%%%%%%%%%%%%%%
One may be interested which of our spaces are parallelizable with respect to
the generalized connection, i.e.
have ${\hat R}^\l_{- \m\n\r}=0$ (and thus ${\hat R}^\l_{+ \m\n\r}=0$, see
\tors).
One expects parallelizable spaces to be related to group spaces
and  indeed this is what we find.

Since the string naturally `feels' the
generalized connection with torsion, the vanishing of the generalized curvature
is the analogue of the flatness condition
in the point-particle theory.
In particular, $\hat R_\pm=0$ means that locally
${\hat \G}^\l_{\pm \m\n}= f^{-1\l}_{\ \pm n} \del_\n f^n_{\pm\m}$.
Then
\conn\   implies the existence of $D$  chiral and $D$ antichiral conserved
currents
$f^n_{-\m} (x) \del x^\m$ and $f^n_{+\m} (x) \bd x^\m$.

 As follows from \cur, a necessary condition for parallelizability
 is
$\del_i\del_j h =0$, i.e. $h=h_0 + p_i x^i$.
Then the two remaining conditions  take the form
\eqn\cooo{{\hat R}^v_{- ijk } =
\del_i{\cal F}_{jk} + 2 p_i {\cal F}_{jk}=0\ ,
} $$ {\hat R}^v_{- iuj}= 2  \del_i\del_u A_j + 4 p_i \del_u A_j
-\del_i\del_j K - 2p_i \del_j K =0\ .   $$
In view of the gauge freedom (2.8)  we may set
 $K=0$.  If $p_i \not=0$ the solution is
$ A_i = C_i (u) \exp (-2 p_j x^j).$
By redefining the coordinates $v'=v + \exp (-2p_i x^i) g(u), \  x'^i = x^i +
w^i(u)$
the corresponding model can be transformed into the  product
of the $SL(2, R)$  WZW model (cf. (B.7))  and $R^{D-3}$.

The case of $p_i=0$, i.e. $F=\const$ is more subtle.
The solution is $ A_i = C_i (u) - \ha {\cal F}_{ij} x^j, \
{\cal F}_{ij}=\const.$
One can further eliminate $C_i$ by a coordinate transformation
$v'=v + q(u) + s_i(u)x^i, \ x'^i= x^i + w^i(u)$.
We are finally left with the following model (cf. \yyys)
\eqn\parr{ F=1\ ,\ \  \ K=0\ ,\ \  \ A_i=-\ha {\cal F}_{ij} x^j \ .}
These   spaces
 can be interpreted as boosted products of group spaces,
or, equivalently, as spaces corresponding to
WZW models for non-semisimple groups. To show this
one should first put ${\cal F}_{ij}$ into the  block-diagonal
form by a coordinate $x^i$ rotation,
so that its elements are represented by constants
$\H_1, ..., \H_{[D/2-1]}$ and the corresponding Lagrangian
(4.10) is (we split $x^i$ into pairs representing 2-planes; $a,b=1,2$)
\eqn\jkl {L= \del u \bd v + \sum_{s=1}^{[D/2-1]}
\(\H_s \ep_{ab} x^a_s \del u \bd x^b_s +
\del x^a_s \bd x_{as} \) \ . }
 The first non-trivial
case is that of $D=4$, i.e. ${\cal F}_{ab} = \H \ep_{ab}$.
The corresponding model
($x_1=r \cos \theta, \ x_2 = r \sin \theta $)
\eqn\moo { L=\del u \bd v + \H \ep_{ab} x^a \bd x^b \del u
+ \del x^a\bd x_a }
$$=\del u \bd v + \H r^2 \bd \theta \del u
+ \del r \bd r + r^2 \del \theta \bd \theta\ , $$
is equivalent to the $E^c_2$ WZW model of ref. \napwi\
(note that $\H$ can be set equal to --1 by a rescaling of
$u,v$).
In fact, the coordinate  transformation \kts\
 $x_1=y_1 + y_2 \cos u   ,
\ x_2 = y_2 \sin u, \ v= v' + y_1 y_2  \sin u $  puts \moo\
in the form
\eqn\fgf{ L = \del u \bd v' + \del y_1 \bd y_1 + \del y_2 \bd y_2
+ \ 2 \cos u\ \del y_1 \bd y_2 \ , }
which is obtained from the  $R\times SU(2)$ WZW action
by a singular boost
and rescaling of the level $k$ or $\a'$ (see \sfts). If
$s$ is a time-like coordinate of the $R$-factor and $\psi$ is an angle
of $SU(2)$ one should set $ s=u, \ \psi = u + \ep v$,
rescale $k$ and $y_i$ by powers of $\ep$ and take $\ep $ to zero.
The $D=5$ model \moo\ is equivalent to the product of the $D=4$
model and a free space-like direction. The $D=6$ model (which contains
two sets of  planar coordinates $x^a_s, \ s=1,2$)
 is
equivalent to the non-semisimple or boosted version
of the $SL(2,R)_{-k_1} \times SU(2)_{k_2}$ WZW model
 (see eq. (4.16) in \sfts).
The required coordinate
transformation is $\psi_1= u , \psi_2 = u + \ep v $, etc.
The non-trivial   parameter $\H_1/\H_2$ is equal
to the ratio of the levels $k_1/k_2$.

The next non-trivial model is with $D=8$.
It can be obtained  by boosting
$SL(2,R)_{-k_1} \times SU(2)_{k_2} \times SU(2)_{k_3}$
 WZW model ($\psi_1 =u, \ \psi_2 = u + \ep v -\ep \l , \
\psi_3 = u + \ep \l$) with the  direction $\l$ decoupling in the limit
$\ep\to 0$.
All higher $D$ models are related to similar WZW models
based on direct products of $SL(2,R)_{-k}, \ SU(2)_k$ and $R$ factors,
 or, equivalently, on corresponding non-semisimple groups.
The parameters $\H_s$  are  essentially equivalent to
the rescaled levels $k_n$ of the factors.

Finally, it is interesting to note that all the models
\parr, like the $D=4$ model \moo, can be related to the
flat space model in the same way  as
 this was shown  \refs{\kiri,\kts}  for the $D=4$ model
of \napwi.
In fact, let us consider one pair of planar coordinates $x^a$
and gauge the rotational symmetry in the plane.
We get the following model
\eqn\modd{ L= \del u \bd v + \del r \bd r + r^2 (\del \theta + A)
(\bd \theta + \bar A)  +
\H r^2 \del u (\bd \theta  + \bar A) + \bar A \del  \tilde\theta -
A \bd \tilde \theta \ , }
where $\tilde \theta$ is the dual coordinate and $A, \bar A$ are
components of the 2d gauge field.
Shifting $A$ by $-\H\del u$
and $v$ by $\H\tilde \theta$ we get a model
which is   equivalent to the flat space one.
The same  transformation can be done independently for each plane.
The original  \sm  \parr\ is  thus related to a flat space one  by a
combination of duality,  coordinate transformation and ``inverse" duality.
If, however, the true starting point is the ``doubled" or ``gauged' model
\modd, then the transformation  to the  model corresponding to the flat space
is just a coordinate transformation on the extended configuration space of
$(u,v,r,\theta,\tilde \theta,A,\bar A)$.

%%%%%%%%%%%%%%%%%%%%%%%%%%%%%%%%%%%%%
\subsec{Leading-order conformal invariance equations}
%%%%%%%%%%%%%%%%%%%%%%%%%%%%%%%%%%%%%
The standard leading-order conformal invariance conditions are
\eqn\lead{  \hat R_{-\m\n} + 2 \hat D_{-\m} \hat D_{-\n} \p =0 \ , }
where $\hat R_{\pm \m\n} =\hat R_{\mp \n\m}= {\hat R}^\l_{\pm\m\l \n} $ and
$\hat D_{\pm\m }$
are the Ricci tensor and covariant derivative  for  the connection $
\hG^\l_{\pm\m\n} $
(the symmetric and antisymmetric parts of \lead\ give equations for $G_{\m\n}$
and $B_{\m\n}$).
Computing the Ricci tensor from \cur\  one finds
\eqn\ric{ \hat R_{-uv}= -  F \del^i \del_i  h  \ ,  \ \ \
 \hat R_{-ij} = - 2 \del_i \del_j h  \ , \ \  \ \hat R_{-iu} =\hat R_{-iv} =
\hat R_{-v\m  } = 0 \ ,
}
%888%
$$
\hat R_{-uu} = -
F (  \ha \del^i \del_i K  + \del^i h \del_i  K +  K \del^i \del_i  h   -
\del^i\del_u A_i  - 2  \del^i h \del_u A_i ) \ ,
$$ $$
\hat R_{-ui}
= -F(\del_j {{{\cal F}^j}_i} + 2 \del_j h {{{\cal F}^j}_i} + 2 A_i  \del^j
\del_j h ) \ .  $$
Then   \lead\ implies
\eqn\leadd{ - \del_i\del_j h + \del_i\del_j \p=0\ , \  \ \ \  \del_i\del_u \p=0
\ ,
\ \ \ \p (x,u)  = \p(u) + b_i x^i  + h (x) \ ,  }
and finally   we get the same relations as in
\qeq,\qeqp
\eqn\qeqq{   -\ha \del^2 F\inv +  b^i \del_i F\inv =0 \ , \ \ \ \  -\ha  \del_i
{\cal F}^{ij} + b_i {\cal F}^{ij} =0 \ , }
\eqn\uuu{
 - \ha  \del^2 K  +  b^i \del_i K +    {\del^i  \del_u A_i} - 2b^i \del_u A_i
+ 2 F^{-1} {\del^2_u \p}  =0\ .}

%%%%%%%%%%%%%%%%%%%%%
\appendix{B}{ General $D=3$ chiral null  model }
%%%%%%%%%%%%%%%%%%%%%
As was shown in \horts\ the generic $D=3$ $F$-model \mof\   is
equivalent to a special $[SL(2,R)\times R]/R$ gauged WZW model
and  can also be identified with the extremal limit of the charged black string
solution of \hoho.
Here we shall consider the generic $D=3$ model
belonging to the class \moxa,
\eqn\moxas{ L_3=F(x) \del u \[\bd v   +
K(x,u)   \bd u + 2 A (x,u)  \bd x \]
+  \del x \bd x     + \a'{\cal R}\p (x,u)\ .   }
  Since the  transverse space here is one-dimensional, one can set $A=0$
  by  a  transformation of $v$ (see \ssa).
  The functions $F, K$ and $ \p$  are then subject to (see \qwqw--\qwqww)
  \eqn\qeqw{   \del^2_x F\inv =2 b \del_x F\inv \ ,
\ \ \ \del^2_x K =  2  b \del_x K
 -4    F^{-1} {\del^2_u \p}\ , \ \ \  \p=  \p(u)  + b x
  + \ha  \ln F(x)\ .    }
Assuming for simplicity that  $K$ and $\p$ do not depend on $u$ we  get the
  following solutions
 \eqn\sss{  F\inv = a + m \e{2bx} \ ,  \ \ \ \   K =   a' +m' \e{2bx}   = c + n
   F\inv (x)  \ ,  }
   so that by shifting $v$ we finish with the following  conformal $D=3$
   model
   \eqn\yyy{  L_3= F(x) \del u \bd v  + n  \del u \bd u +
    \del x \bd x  +  \a'{\cal R} \p (x) \ , }
    \eqn\ww{ F\inv = a + m \e{2bx} \ , \ \ \ \   \p(x) =\p_0 - \ha \ln
(a\e{-2bx} +
    m)  \ . }
    $a, n, m $ are  arbitrary constants which  take only   two non-trivial
values: 0
      and 1  (--1 case is related to the +1  one by an analytic continuation).
      In what follows we shall set $m=1$.
      The $n=0$ model is   the \FM discussed in \horts.
In what follows we shall
keep $n$ general thus  treating  both $n=0$ and $n=1$ cases at the same time.

The solution \sss\ with  $a=0$  has  a constant dilaton  and thus the
corresponding
model must be
equivalent to the $SL(2,R)$  WZW model (since there are no other $\p=\const$
solutions in $D=3$ in a properly chosen scheme \horts).
In fact, the $ a=0$ model
\eqn\yyyy{  L_3=  \e{-2bx }   \del u \bd v  + n  \del u \bd u +
 \del x \bd x  +  \a'{\cal R} \p_0 \ , }
 is related to the   $SL(2,R)$ WZW Lagrangian written in the Gauss
decomposition
 parametrization (we follow the notation of \horts\  and  set $\a'=1$)
 \eqn\yyz{  L_{wzw} = k \( \e{-2r }   \del u \bd v  +
  \del r \bd r \) \  , }
by the following coordinate transformation ($u',v'$ stand for the  coordinates
 in \yyz)
  \eqn\coor{
   u'={1\ov 2 \sqrt n}  \e{2b\sqrt n u} \ , \ \ \
   v'= bv -  \sqrt n \e{ 2 bx} \ , \ \ \
   r = bx + b \sqrt n   u  \ , \
   \ \   b^2 = 1/k \ . }
   %%%%%%%%%%%%%%%%%%%%%%%%%%%%%%%%%%%%%%
   \subsec{Gauged WZW model interpretation}
   %%%%%%%%%%%%%%%%%%%%%%%%%%%%%%%%%%%%%%%%
Like the $n=0$ model,  the $n=1$ one \yyyy\  can be  related to a    special
$[SL(2,R)
    \times R]/R$ gauged WZW model.
     This  provides an  explicit illustration of   our claim that
     the chiral null backgrounds   are exact conformal models.\foot{It is very
likely that there exists a generalization of the nilpotent gauging
procedure of ref. \klts\  which    makes it possible to identify not just one
$D=3$ model
 but a whole subclass of the  chiral null models  with $  F\inv =  {
\sum_{i=1}
 ^d  {\rm e}^{  \a_i\cdot x  } }\  $ ($\a_i$
 are the simple roots of the   algebra of a maximally non-compact
 Lie group  of rank $d=D-2 $)  with the gauged WZW models.}
The $SL(2,R)\times R$ WZW model  written in the Gauss decomposition
parametrization, i.e. \yyz\ with an additional $R$-term $ \   \del y \bd y $,
 has the following obvious global symmetries:  independent shifts of $u,v,y$
and
 shifts of $r$ combined with rescalings of $u$ and (or)  $v$.
 Gauging the  translational subgroup
 \eqn\fff{  u\ra   u + \ep \  , \ \  \ v\ra v+ \ep\  , \ \ \  y\ra y + \r \ep\
, \ \ \ \r=\const\ , }
 fixing the gauge $y=0$ and integrating out the two dimensional
 gauge field,  one gets the $n=0
 $ model \yyy\  with $a=\r^{-2}$ \horts.
 The subgroup which is to be gauged to   get the $n=1$ model  is\foot{It  may
be useful to  recall that the subgroup that  leads to  the charged
black string of \hoho\ is \horts:
$u\ra \e{\ep} u, \   \ v\ra \e{\ep} v ,  \  r\ra r+ \ep , \  y\ra y + \r \ep
.$}
 \eqn\suu{ u\ra \e{2\sqrt n\ep} u \ , \ \  \ v\ra v+ \ep\ , \ \  \  r\ra r+
\sqrt
  n \ep\ , \ \ \  y\ra y + \r \ep\ .   }
  In view of \coor\ this is just the translational symmetry \fff
  \ (with $\ep \ra b\inv \ep$)
  of the action \yyyy\  (with $   \del y \bd y $ added).
  Since \yyyy\ is  a coordinate transformation of the WZW action
  \yyz\ we can start   directly with  \yyyy\  in  the gauging procedure,
   \eqn\wwss{   L_{gwzw}  =
      {\rm e}^{-2bx}  (\del u  + A) (\bd v  + \bar A  )  +  n  (\del u  + A)
(\bd u  +
      \bar A  ) + \del x \bd x
      +   (\del y +  \r A ) (\bd y +  \r \bar A ) \ .  }
      Fixing $y=0$ as a gauge  and integrating out $A,\A$ we get
      \eqn\wws{   L_{gwzw}  =
       { \r^2  {\rm e}^{-2bx} \ov \r^2 + n +  {\rm e}^{-2bx}  }
	\del u  \bd v + { n \r^2\ov  \r^2 + n +  {\rm e}^{-2bx} } \del u  \bd u } $$
	  +\  \del x \bd x
  +    \a'{\cal R} [ \p_0'  + \ha \ln ( \r^2 + n +  {\rm e}^{-2bx})] \ .  $$
	  The  redefinition
	  \eqn\rede{u'= (1 + n\r^{-2} )^{1/2} u      \ , \ \ \ \
	       v'=   (1 + n\r^{-2} )^{1/2} (v + {n\ov \r^2 + n }u)   \ , }
puts this action into the desired form \yyy,\ww\ with  $ a= (\r^2 + n )^{-1}$.

%%%%%%%%%%%%%%%%%%%%%%%%%%%%%%%%%%%%%%%%%%%%%%%%%
\subsec{Extremal black string interpretation}
%%%%%%%%%%%%%%%%%%%%%%%%%%%%%%%%%%%%%%%%%%%%%%%%%

The generic $D=3$ $F$-model (i.e. \yyy\ with $n=0$)  can be considered as an
extremal limit of the charged black string solution of \hoho.
Here we point out that a similar statement is true for the $n=1$ model \yyy.
This is  a particular case of the relation between  the model \mofke\
and the charged black string solution discussed in Section 2.4 (see \bksth).
 Starting with the non-extremal charged black string \sm which has
 the metric
 \eqn\ppp{ ds^2 = -  f_1 (r')  dt'^2 + f_2 (r')  dy'^2  +  h(r')  dr'^2
  \ ,  } \eqn\hhh{
f_1= (1- {M_1\ov r'})\ ,  \ \ f_2= (1- {M_2\ov r'}) \ , \ \
\ h=(4 r'^2 f_1 f_2)\inv\ ,  \ \  \  M_1=M\ , \ \ M_2={ Q^2\ov M} \ , }
  boosting the solution
\eqn\iii{t= \l v + (\ha \l - \l\inv) u \ , \ \  \ y= \l\inv u \ , \ \ \
\l\equiv  ({M_1\ov M_
   2 }-1)^{1/2} \ ,  }
and then taking the extremal limit $M\ra Q, $ i.e.  $M_1\ra M_2$ or $ \l\ra 0$
   in the resulting \sm
   one finishes  with the model \yyy\ with the metric
\eqn\tty{ ds^2 = 2(1-{M\ov r'}) dudv +  du^2 +  h(r') dr'^2 =  F(x) dudv + du^2
+ dx^2 \  . }
So the generic $u$-independent $D=3$ chiral null model can be obtained
as an extremal limit of a black string solution.

\vfill\eject

\listrefs
\end